\documentclass[11pt]{article}

\usepackage[letterpaper,margin=1.4in]{geometry}
\usepackage{cite}
\usepackage{amsmath,amssymb,amsfonts}
\usepackage{booktabs}
\usepackage{array}           
\usepackage{amsthm}
\theoremstyle{definition}   
\newtheorem{definition}{Definition}

\usepackage{graphicx}
\usepackage{textcomp}
\usepackage{xcolor}
\usepackage{microtype}
\usepackage{authblk}

\usepackage{algorithm}
\usepackage[noend]{algpseudocode}
\usepackage{multirow}
\usepackage{pgf-pie}   
\usepackage{pifont}
\usepackage{pgfplots}
\usetikzlibrary{patterns}
\pgfplotsset{compat=1.18}

\usepackage[hidelinks]{hyperref}

\newcommand{\figonewidth}{0.78\textwidth}   
\newcommand{\figtwowidth}{0.72\textwidth}   
\newlength{\craftfigbox}                     
\newlength{\figfourbox}
\newcommand{\figfourboxa}{2.50in}
\newcommand{\figfourboxb}{4.05in}
\newcommand{\figfourfont}{\scriptsize}   
\newcommand{\figfourwrap}[1]{\makebox[\linewidth][c]{#1}}
\providecommand{\Description}[1]{} 
\newcommand{\artifactnote}{} 

\makeatletter
\renewenvironment{figure*}{\@float{figure}}{\end@float}
\renewenvironment{table*}{\@float{table}}{\end@float}
\makeatother

\title{\Large\textbf{CRAFT: Constrained Route planning with Adaptive\\
Framework and Template-based code generation}}

\author{Shahd Elmahallawy}
\author{Ahmed Eldawy}
\affil{University of California, Riverside\\ \texttt{selma008@ucr.edu} \quad \texttt{eldawy@ucr.edu}}
\date{}

\begin{document}
\maketitle

\begin{abstract}
Everyday routing requests are rarely a point-to-point shortest path. A driver may need to visit stops, sometimes in a fixed order, under time budgets or distance limits, and perhaps recharge their EV car along the way. Each combination is a distinct routing problem that classically requires its own algorithm, so no fixed solver covers the space of queries users may ask. This paper presents CRAFT, which takes a constrained routing query as free text and returns a route that satisfies the user's requirements on a real road network. An LLM parses the query and generates four search functions, which a generic graph-search engine executes on the road network. We formalize the underlying problem as State-Expanded RCSP (SE-RCSP), which extends resource-constrained shortest paths (RCSP) with non-additive state such as node-dependent resets and discrete progress tracking. We evaluate CRAFT on 2,600 natural-language queries over California road networks against two baselines, analyzing correctness, constraint violations, long-distance scalability, and semantic POI filtering.
\end{abstract}

\noindent\textbf{Keywords:} constrained route planning, large language models


\providecommand{\craftfigbox}{\columnwidth}

\section{Introduction}
\label{sec:intro}
Online maps serve billions of users and answer massive volumes of routing queries every day~\cite{ZZLZ25}. These systems build on the strict definition of the shortest/fastest path problem, which allows heavy precomputation and answers in milliseconds~\cite{bast2016route,S14}. Many practical trips, however, are not point-to-point shortest paths. They are constrained route-planning tasks in which users need to satisfy semantic, temporal, resource, and ordering constraints while still optimizing travel cost such as for travel planning~\cite{aaa}. Constraints of the numeric kind have a classical formulation, the resource-constrained shortest path problem (RCSP)~\cite{pugliese2013survey,irnich2005spprc}, which minimizes path cost while the quantities consumed along the route, such as travel time or tolls, stay within their limits; recent algorithms such as ERCA*~\cite{ren2023erca,ahmadi2025fast}, together with index-based methods for multi-constraint~\cite{liu2023forest} and label-constrained~\cite{zhang2022lcsp} variants, make it practical on road networks. What users ask for, though, is beyond a simple predefined numeric budget. Take the query in~\autoref{fig:toy_example}: a user at $v_1$ drives an electric vehicle with 10\% battery and wants to stop by a \emph{park with a playground} and then a \emph{restaurant} before reaching $v_{10}$. The correct answer, highlighted in the figure, first detours to a charging station at $v_4$, then visits the park at $v_7$ and the restaurant at $v_9$.

Queries like this span several problem families, and~\autoref{fig:toy_example} moves through them one requirement at a time. The two stops, taken in the requested order, form the optimal sequenced route query~\cite{sharifzadeh2008osr,levin2010interactive,liu2018topk}, and a route search query when the order is left free~\cite{li2005trip,cao2012kor}. Another line of work targets queries that name no stops at all and ask instead to visit as many places as possible before a deadline: this is itinerary planning, rooted in the orienteering problem~\cite{vansteenwegen2011orienteering,bolzoni2014,lu2012personalized,vathis2024regular}. None of these three carries any notion of a resource, so the battery brings in a fourth family, electric-vehicle routing~\cite{baum2019chargingstops,storandt2012cruising,LZW16}, and it also puts the query beyond RCSP, which tracks only quantities that are spent and never restored. Each of these families is served by dedicated algorithms over structured input, and, to the best of our knowledge, no existing solver accepts an arbitrary combination of them, even though the combination first appears in the user's query: the trip in~\autoref{fig:toy_example} already needs a battery constraint and ordered stops at once.

\providecommand{\figonewidth}{0.50\textwidth}
\begin{figure}[t]
    \centering
    \includegraphics[width=\figonewidth]{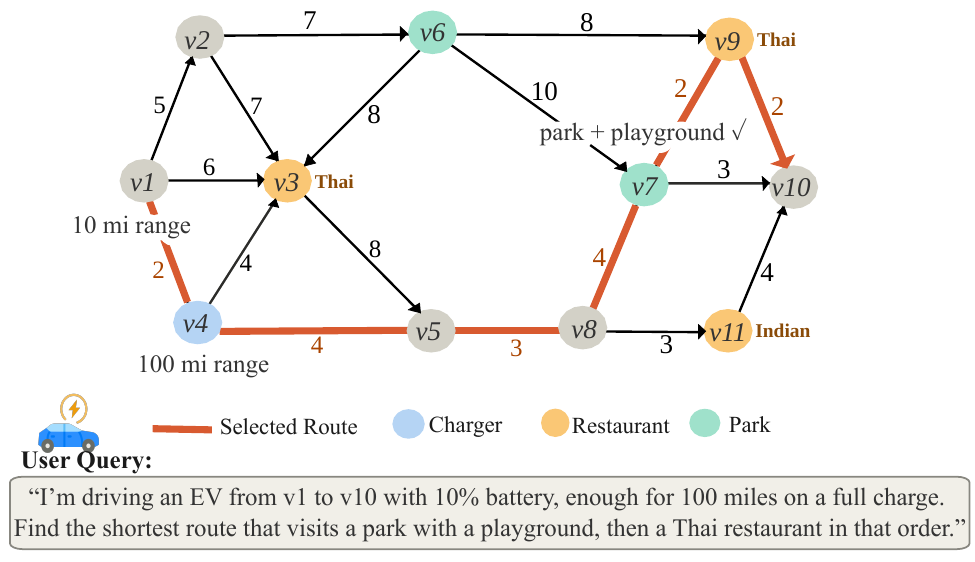}
    \caption{User Query Example}
    \label{fig:toy_example}
\end{figure}

Large language models (LLMs) offer that interface, but the LLM alone is a poor planner. On a benchmark of multi-constraint travel planning, GPT-4 succeeds on less than 1\% of tasks, hallucinating places and fabricating distances so that its plans violate the stated constraints~\cite{xie2024travelplanner,shao2026chinatravel}. Recent systems therefore use the LLM only to interpret the query in front of one fixed solver, e.g., SpeakNav matches spoken descriptions to a template-driven path search~\cite{zheng2021speaknav}, ITINERA feeds an itinerary optimizer~\cite{tang2024itinera,pugliese2025spatially}, and LLMAP parses stops for a multi-step graph search~\cite{yuan2025llmap}. This restores reliability, since the route comes from an algorithm rather than from the model, but the interface still covers only the constraint types that solver was built for. A separate line of work lets the LLM write the planning logic itself, but for different problems. The Automatic Routing Solver (ARS) generates code for vehicle routing, which assigns a set of customers to a fleet of vehicles rather than searching for a path between two points~\cite{li2025ars}. Hao et al.\ have the LLM rewrite a travel request as a list of formal conditions, then use an automated tool to check whether a proposed multi-day plan meets them, without any road network~\cite{hao2025large}. Both show that an LLM can turn a planning problem into executable form, which is the idea we build on, but neither answers a constrained routing query on a road graph.
 
This paper introduces CRAFT\artifactnote{}, an adaptive constrained routing framework that answers natural-language route-planning queries on road networks. Instead of one solver per constraint family, CRAFT defines the \emph{state-expanded resource-constrained shortest path} (SE-RCSP), in which each classical family and any combination of them is an instance specified by four functions. Instead of a structured input format, the Spec Builder turns the free-text query into a structured form aligned with the road network and POI schema, and the ORCStrator uses an LLM to write the four functions for that query, so coverage is not bounded by pre-built algorithms. The generated functions plug into a robust SE-RCSP Solver that extends Dijkstra's algorithm and runs on the real graph, so the model never produces the route itself.
 
We evaluate CRAFT on 2,600 natural-language queries over California road networks, 1,600 of which we introduce and another 1,000 inspired by an existing benchmark~\cite{yuan2025llmap}, and compare against LLMAP~\cite{yuan2025llmap} and an LLM-only baseline across three LLM backends. CRAFT answers up to 93\% of the main benchmark correctly, against at most 42\% for the strongest baseline, and returns the best route on up to 88\% of the queries.
 
To summarize, this paper makes the following contributions:
\begin{enumerate}
    \item We formally define the \textbf{SE-RCSP} problem for flexible constrained shortest path.
    \item We introduce CRAFT, a framework that uses LLM to translate a natural-language routing query into an executable instance of SE-RCSP.
    \item We propose a robust graph search algorithm to solve an SE-RCSP problem instance on a real road network and POI database.
    \item We build an extensive benchmark of 2,600 natural-language queries and evaluate the correctness and performance of CRAFT compared to two LLM baselines.
\end{enumerate}
 
The remainder of the paper is organized as follows. Section~\ref{sec:related_Work} reviews related work. Section~\ref{sec:problem} formulates SE-RCSP. Section~\ref{sec:system} gives a system overview, and Sections~\ref{sec:input-parser} to~\ref{sec:generator} describe the Spec Builder, the ORCStrator, and the SE-RCSP Solver. Sections~\ref{sec:benchmark} and~\ref{sec:experiments} present our benchmark and experimental results. Section~\ref{sec:conclusion} concludes the paper.

\section{Related Work}
\label{sec:related_Work}
This section summarizes the related work in four areas. First, it covers three classes of planning problems, \emph{route planning}, \emph{itinerary planning}, and \emph{trip planning}. Then, it discusses the broad problem of using LLM for query parsing and code generation.
The first two problems are the closest to this paper and are further summarized in \autoref{tab:taxonomy} and classified based on the objectives and constraints they support. 

 
\begin{table}[t]
\centering
\caption{Related work on route planning}
\label{tab:taxonomy}
\footnotesize
\setlength{\tabcolsep}{3.5pt}
\begin{tabular}{@{}lccccl@{}}
\toprule
 & \multicolumn{2}{c}{\textbf{POIs}} & \multicolumn{2}{c}{\textbf{Route cost}} & \\
\cmidrule(lr){2-3}\cmidrule(lr){4-5}
\textbf{Problem} & \textbf{Obj.} & \textbf{Constr.} & \textbf{Obj.} & \textbf{Constr.} & \textbf{Related work} \\
\midrule
Shortest path        &            &            & \checkmark &            & \cite{bast2016route, delling2017crp} \\
RCSP                 &            &            & \checkmark & \checkmark & \cite{pugliese2013survey, ren2023erca, ahmadi2025fast, cola, liu2023forest, zhang2022lcsp, wang2024pcsp} \\
Routing with stops   &            & \checkmark & \checkmark & (\checkmark) & \cite{li2005trip, sharifzadeh2008osr, liu2018topk, zhu2022rcosr} \\
Itinerary planning   & \checkmark & (\checkmark) &          & \checkmark & \cite{vansteenwegen2011orienteering, bolzoni2014, vathis2024regular} \\
\midrule
CRAFT                & \checkmark & \checkmark & \checkmark & \checkmark & this paper \\
\bottomrule
\end{tabular}

\begin{flushleft}
(\checkmark) Supported only by some of the cited work.
\end{flushleft}
\vspace{-15pt}
\end{table}

\subsection{Route Planning}
Route planning problems are designed to minimize the route cost, e.g., distance or time, and may include additional constraints.
We further divide route planning into three categories as follows:

\textbf{I. Shortest path.} Classical algorithms such as Dijkstra's and A$^\star$ form the backbone of road-network routing, with speedup techniques such as contraction hierarchies~\cite{bast2016route,delling2017crp}. They optimize a single objective, typically distance or travel time, with no constraints along the path.

\textbf{II. Constraints on the edges.} The resource-constrained shortest path (RCSP) problem adds limits on quantities that accumulate as edges are traversed, such as total travel time, toll cost, or energy consumed, and finds the cheapest path that keeps each within its limit~\cite{pugliese2013survey}. Solvers that adopt this formulation, e.g., ERCA*~\cite{ren2023erca} and NWRCA*~\cite{ahmadi2025fast}, use label-setting search with dominance pruning. To further optimize these techniques, some techniques precompute an index for one fixed constraint type and answers such queries by lookup, for example a budget on an accumulated resource~\cite{cola, liu2023forest} or road categories to avoid~\cite{zhang2022lcsp, wang2024pcsp}.

\textbf{III. Constraints at the nodes.} In \emph{sequenced route queries}, the user specifies specific POI categories, and the search finds a route with concrete POIs to serve them. The order may be fixed~\cite{sharifzadeh2008osr, liu2018topk, zhu2022rcosr, zheng2021speaknav} or free~\cite{li2005trip}.

CRAFT builds on these problem formulations but generalizes to support a combination of constraints that goes beyond the ones described above.

\subsection{Itinerary Planning and Orienteering}
Itinerary planning inverts the objective by using the visited POIs as the objective and the travel time/distance as a constraint. Its base formulation is the orienteering problem: each POI has a score, and the goal is to collect the maximum total score within a travel budget~\cite{vansteenwegen2011orienteering}. Stops are optional, so leaving one out still yields a feasible plan with a lower score. Later work extends it with additional constraints: Bolzoni et al.~\cite{bolzoni2014} limit how many POIs of each category may be visited, and Vathis et al.~\cite{vathis2024regular} require the visited categories to follow a given pattern and respect opening hours. What separates this family from routing is that stops are optional and the objective maximizes reward rather than minimizing cost: routing through required stops is TSP-like, whereas selecting optional POIs under a budget is knapsack-like~\cite{zhao2024semantic}.

CRAFT supports the orienteering problem and uses optimization techniques to speed up its execution, as detailed later.

\subsection{Travel and Trip Planning}
Travel planning schedules bookable items, such as lodging, dining, and transport, into a multi-day plan. Benchmarks such as TravelPlanner~\cite{xie2024travelplanner} and ChinaTravel~\cite{shao2026chinatravel} formalize this setting, and follow-up work extends them~\cite{flextravelplanner, triptailor, compass, tprag, tripscore}. The road network disappears in this formulation as it uses a matrix to lookup travel time between items rather than a computed path, so a system can score perfectly without ever producing a route. This is why we do not evaluate on these benchmarks.

\subsection{LLMs as Parsers and Code Generators}
Using an LLM as a direct planner was found to be unreliable. On TravelPlanner, GPT-4 reaches only a 0.6\% success rate~\cite{xie2024travelplanner}. Two more robust roles have emerged. As a \emph{parser}, the LLM turns the request into a structured form for a separate backend, as in ITINERA~\cite{tang2024itinera}, LLMAP~\cite{yuan2025llmap}, and Pugliese et al.~\cite{pugliese2025spatially}. The backend still supports a fixed set of constraints, so the LLM widens the interface but not the problem coverage. As a \emph{code generator}, it writes the logic as executable code, as in ARS~\cite{li2025ars} and Hao et al.~\cite{hao2025large}. CRAFT further investigates the reasons of failure for LLM as a planner and proposes a more efficient framework for using LLM as both a parser and code generator while fixing a robust code executor to ensure correctness and efficiency.

\section{Problem Formulation}
\label{sec:problem}
Given a road network and a set of points of interest, a user needs a route from a source location to a destination that meets some requirements along the way. In \autoref{fig:toy_example}, the user drives from $v_1$ to $v_{10}$ with 10\% battery and must stop at a park with a playground and then at a Thai restaurant. The correct route detours to a charging station at $v_4$, then visits the park at $v_7$ and the restaurant at $v_9$. This section formalizes the problem, from the road network to the route that answers the query.

\subsection{Preliminaries}
\begin{definition}[Road Network]
\label{def:graph}
A road network is a directed graph $G=(V,E)$. A node $v \in V$ is an intersection with a geographic location $\mathit{loc}(v)$. An edge $(u,v) \in E$ is a road segment from $u$ to $v$ with a set of attributes $\mathit{attr}(u,v)$, such as its length $\ell(u,v)$, travel time $\tau(u,v)$, and road class.
\end{definition}

Each node $v$ hosts a set of POIs $\mathit{POI}(v)$, possibly empty. In this paper, we associate each POI with its nearest road intersection, but any other approach may be used. We use the notation $(u,v)$ to represent the \emph{edge structure} that contains the attributes of both end vertices and the edge.

\begin{definition}[Point of Interest]
\label{def:poi}
A point of interest (POI) is a place $p=(\kappa, A, d)$ with a category $\kappa$ (e.g., restaurant), a set of structured attributes $A$ (e.g., cuisine), and an optional text description $d$ (e.g., ``has a playground''). A POI hosted at a node $v$ takes the location of that node, $\mathit{loc}(v)$.
\end{definition}

\begin{definition}[Path]
\label{def:path}
A path is a sequence of nodes:
\begin{equation}
P=\langle u_0, u_1, \ldots, u_k\rangle,
\end{equation}
such that $u_i\in V$, $(u_i,u_{i+1})\in E$ for every $0 \le i < k$.
\end{definition}

\subsection{State}
A path specifies only the roads driven. The user's requirements, however, concern quantities that change along it, such as the battery left, the visited POIs, or the elapsed time. Following the RCSP literature~\cite{irnich2005spprc}, we call these quantities \emph{resources} and track them along the path.

\begin{definition}[Resource]
\label{def:resource}
A resource $R_j$ is a variable that tracks a quantity along the path. It can be any value the query needs to track, numeric or categorical, such as the battery left, the elapsed time, or the set of stops visited so far. The resources of a query form the vector $\mathbf{R}=(R_1,\ldots,R_m)$. When the path traverses an edge $(u,v)$, each resource is updated to a new value
\begin{equation}
\mathbf{R}' = g(\mathbf{R}, (u,v)),
\label{eqn:resource}
\end{equation}
where $g$ is a function that computes the updated resource vector $\mathbf{R}'$ based on the initial resource vector $\mathbf{R}$ and the edge $(u,v)$. The function $g$ may read the current resources, the attributes of the edge, and the attributes of its endpoints, including the POIs hosted at the node reached.


\end{definition}
In the example in~\autoref{fig:toy_example}, the resource vector includes the number of required POIs visited and the remaining battery.

\begin{definition}[Objective]
\label{def:objective}
The objective is another numeric value accumulated along the path. It starts at $0$ at the source and, at each edge, adds a step cost using a function $\delta$:
\begin{equation}
f_0 = 0, \qquad f_{i+1} = f_i + \delta\bigl(\mathbf{R}_{i+1},(u_i,u_{i+1})\bigr),
\label{eqn:objective}
\end{equation}
where $\mathbf{R}_{i+1}$ is the resource vector after the edge is traversed. The query defines the step cost, $\delta$, based on the edge traversed, e.g., travel time, or the node reached, e.g., dwell time at POI. The objective of the path is the objective of the destination node, i.e., $f(P)=f_k$, and the query may ask to minimize or maximize it. We treat a maximization problem as minimization of $-f$, so we assume minimization from here on.
\end{definition}

\begin{definition}[State]
\label{def:state}
A state at node $v$ is a tuple
\begin{equation}
s=(v,\ \mathbf{R},\ f),
\end{equation}
where $\mathbf{R}$ is the resource vector at $v$ and $f$ is the objective accumulated from the source to $v$.
\end{definition}
The objective in~\autoref{fig:toy_example} is the total path length.

\begin{definition}[State-Expanded Path]
\label{def:sepath}
A state-expanded path is a path $P=\langle u_0,\ldots,u_k\rangle$ together with the states $\langle s_0,\ldots,s_k\rangle$ it induces, where $s_0$ is the initial state given as part of the input, $s_{i+1}$ is computed from $s_i$ using Equations~\ref{eqn:resource} and~\ref{eqn:objective}.
\end{definition}

\subsection{Feasible Path}
Not every path satisfies the user query. User requirements are conditions on the resources the path carries, e.g., the battery is a resource, and keeping the charge above zero is a requirement. We formulate such requirements as constraints, and a path that meets all of them is called a \emph{feasible path}.

\begin{definition}[Constraint]
\label{def:constraint}
A constraint is a predicate $\varphi(\mathbf{R}) \in \{\mathrm{true},\mathrm{false}\}$ on the resource vector.
\end{definition}

\smallskip
For example, $\varphi(\mathbf{R}) = (R_{\text{battery}} > 0)$ requires that the battery never runs out. Each query defines its own set of constraints $C = \{\varphi_1, \ldots, \varphi_n\}$. Neither the number $n$ nor the form of each $\varphi_i$ is fixed. In other words, any requirement that can be checked from the tracked resources is a valid constraint.

We classify constraints into \emph{step} and \emph{trip} constraints. A step constraint $C_{\text{step}}$ must hold at every state along the path, e.g., the battery limit above. The trip constraint $C_\text{trip}$ must hold at the destination, such as having visited all required stops.

\begin{definition}[Feasible Path]
\label{def:feasible}
A state-expanded path $P=\langle u_0,\ldots,u_k\rangle$ carries a resource vector $\mathbf{R}_i$ at each node $u_i$, obtained by applying $g$ along its edges. $P$ is feasible if $u_0=v_s$, $u_k=v_d$ (the query source and destination), every step constraint holds at every node, and every trip constraint holds at the destination:
\begin{equation}
\begin{aligned}
\varphi(\mathbf{R}_i) &= \mathrm{true} && \forall \varphi \in C_{\mathrm{step}},\ 0 \le i \le k, \\
\varphi(\mathbf{R}_k) &= \mathrm{true} && \forall \varphi \in C_{\mathrm{trip}}.
\end{aligned}
\label{eqn:feasible}
\end{equation}
\end{definition}

\subsection{SE-RCSP Problem}
\label{sec:sercsp-problem}
A query may have many feasible paths. To choose among them, we compare them by the objective of Definition~\ref{def:objective}.

\begin{definition}[Query]
\label{def:query}
A query is a tuple
\begin{equation}
Q=(v_s,\ v_d,\ R_0, g, \delta, C_\text{step},C_\text{trip}),
\end{equation}
where $v_s$ and $v_d$ are the source and destination nodes, $R_0$ is the initial resource vector at $v_s$, $g$ and $\delta$ are the resource and objective update functions, respectively, and $C_\text{step}$ and $C_\text{trip}$ are the two sets of constraints.
\end{definition}


\begin{definition}[SE-RCSP]
\label{def:sercsp}
Given a road network $G$ and a query $Q$, the \emph{State-Expanded Resource-Constrained Shortest Path} (SE-RCSP) problem finds a cost-optimal feasible path from $v_s$ to $v_d$, i.e., satisfies all the constraints and has the smallest objective:
\begin{equation}
P^{*}=\operatorname*{arg\,min}_{P \in \mathit{FP}(Q)} f(P),
\end{equation}
where $\mathit{FP}(Q)$ is the set of feasible paths of $Q$. If $\mathit{FP}(Q)$ is empty, the query is infeasible.
\end{definition}

This formulation covers both query families that CRAFT targets, route planning and itinerary planning, through the choice of resources and objective. 
\section{System Overview}
\label{sec:system}
\begin{figure}[t]
    \centering
    \providecommand{\figtwowidth}{1.0\linewidth}%
    \includegraphics[width=\figtwowidth]{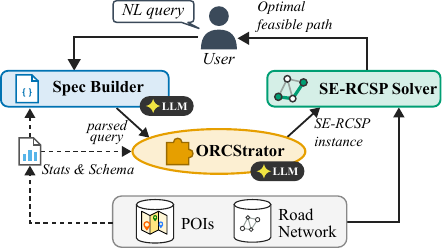}
    \caption{CRAFT System Overview}
    \Description{This diagram illustrates the CRAFT pipeline.}
    \label{fig:system}
\end{figure}

This section gives an overview of CRAFT, the framework that turns a natural-language query into an SE-RCSP instance and solves it.
Figure~\ref{fig:system} shows the three components, which we illustrate with the example query from Figure~\ref{fig:toy_example}.

\smallskip
\noindent\textbf{Spec Builder}
(\autoref{sec:input-parser}) converts the natural-language query into a structured specification in three steps: (1)~\emph{POI Schema Loading} reads the POI categories, attributes, and road classes in the dataset, so the LLM only references what exists in it. This is helpful when the user's query is vague and doesn't use same categories in the dataset. (2)~\emph{Query Extraction} uses the LLM to identify the origin, destination, objective, and all constraints; and (3)~\emph{Resolution and Normalization} maps coordinates to graph nodes and standardizes units, e.g., battery range in meters.

\smallskip
\noindent\textbf{SE-RCSP Solver}
(\autoref{sec:engine}) is a single, query-agnostic search engine that solves every SE-RCSP instance on the real road network. It employs a best-first search over states and returns a best feasible route for any combination of constraints, without being rewritten for each new query type.
Each query reaches the solver as four functions over states (\emph{Core Functions}) that initialize the state, update it along an edge, compare two states, and recognize a feasible route, which together capture the query's objective, resources, and constraints. The solver also prunes states that cannot lead to a better route (\emph{Optimization}). For the example, it returns a route from $v_1$ to $v_{10}$ through a charging station, a park with a playground, and a Thai restaurant.

\smallskip
\noindent\textbf{ORCStrator}
(\autoref{sec:generator}) The \textbf{O}bjective, \textbf{R}esources, \textbf{C}onstraints, and \textbf{St}ate gene\textbf{rator} connects the other two components by formulating an SE-RCSP query instance as four functions that carry all query logic and are ready to be executed by the SE-RCSP Solver. It first assembles a prompt from the structured query and the schema (\emph{Prompt Assembly}), and then uses the LLM to write the four bodies into a fixed template (\emph{Filling the Template}). It also matches free-text descriptions against the POIs as the search reaches them (\emph{Matching Stops}), and finally hands the filled template to the SE-RCSP Solver, which drives the search.
\section{Spec Builder}
\label{sec:input-parser}
The purpose of this step is to convert the free-text user query into a structured query to prepare it for the routing algorithm. This reduces hallucination, since the model only works with values that exist in the dataset. The SE-RCSP query of \autoref{sec:problem} must be written against concrete values, e.g., node identifiers, dataset categories, and limits in the units that the graph uses. A free-text query may provide none of these, so the \emph{Spec Builder} produces them in three steps: \textbf{schema loading}, which tells the LLM what the dataset actually contains, \textbf{query extraction}, which turns the request into a structured query, and \textbf{resolution and normalization}, which aligns it with the graph and its units. \autoref{fig:parser-output} shows the parsed query for the example in \autoref{fig:toy_example}.

\smallskip
\noindent\textbf{Step 1: POI Schema Loading.}
If we ask an LLM for ``a Thai restaurant'', it might invent a category name that seems plausible, or a restaurant name that is not in our dataset, yielding no results. Therefore, this step loads the dataset schema first, including POI categories, other POI attributes, e.g., cuisine and wheelchair accessibility, as well as edge attributes such as road classes. It then computes statistics for these attributes, including all unique values, the range of numeric attributes, and sample values to determine their formats.
The schema and stats are appended to the prompt in the next step, so the model maps requirements onto names that exist (``a Thai restaurant'' $\rightarrow$ poi.cat = `restaurant' $\wedge$ poi.cuisine = `thai').

\smallskip
\noindent\textbf{Step 2: Query Extraction.}
Query extraction takes the first step towards creating a structured query.
Given the natural language query and the stats, the LLM emits a structured query with four parts: (i)~\textbf{source and destination}, copied as written, e.g., address or landmark name;
(ii)~the \textbf{objective}, either minimizing distance or travel time, or maximizing the number of POIs visited;
(iii)~the \textbf{constraints}, such as POI categories, visit order, time budget, per-leg distance limit, or battery range; and
(iv)~\textbf{optional POI filters} that go beyond a category. These are index-aligned with the POI types, each either null or a qualitative requirement. For example, in \autoref{fig:parser-output} each of the two POIs has an additional filter based on the query.
Numeric constraints are kept as value--unit pairs in the units the user wrote, e.g., `100 miles', `10\%'. In \autoref{fig:parser-output}, \emph{park} carries the filter \emph{playground for kids} while \emph{restaurant} carries \emph{cuisine=`thai'}. The types and filters are what decide which POIs can serve each stop, checked as the search reaches them (Section~\ref{sec:filtering}).

\smallskip
\noindent\textbf{Step 3: Resolution and Normalization.}
This final step converts the parsed values into the \emph{units} that the generated code can directly use during query execution.
The generated code will access all attribute values directly from the graph; a mismatch in units would result in an incorrect answer rather than raising an error. This step removes that risk by matching the parsed query parts with the same units used in the dataset. Source and destination are geolocated and snapped to the nearest node by haversine distance, anchoring them to routable positions. Every value--unit pair is converted to the units of the data, e.g., distances to meters and times to seconds, and a battery given as a percentage becomes an absolute range in meters using the vehicle range in the query. If a time budget is present but no per-stop duration is given, each stop is assumed to take 30 minutes.

\begin{figure}[tb]
\centering
\footnotesize
\setlength{\tabcolsep}{3pt}
\begin{tabular}{@{}>{\raggedright\arraybackslash}p{0.20\columnwidth}>{\raggedright\arraybackslash}p{0.8\columnwidth}@{}}
\toprule
\textbf{Field} & \textbf{Value} \\
\midrule
query\_id      & example\_query\_1 \\
source         & v1 \\
destination    & v10 \\
objective      & minimize\_distance \\
poi\_category  & [park, restaurant] \\
poi\_filters   & [\{description: `playground for kids'\}, \{cuisine: `thai'\}] \\
ordered        & true \\
max\_battery   & 160934.0 (meters) \\
current\_battery & 16093.4 (meters) \\
\bottomrule
\end{tabular}

\caption{Spec Builder output for the example in~\autoref{fig:toy_example}}
\label{fig:parser-output}
\end{figure}

The output of the Spec Builder is a structured JSON object similar to the one in \autoref{fig:parser-output}, \emph{source} and \emph{destination} as node IDs, \emph{objective}, \emph{poi\_categories} and \emph{poi\_filters} for the required stops and their qualitative requirements, \emph{ordered} for whether the sequence is fixed, and the remaining fields for the normalized numeric constraints, here the battery state, in the units the SE-RCSP Solver expects.


\section{SE-RCSP Solver}
\label{sec:engine}

The solver computes the route that answers a parsed query by solving the SE-RCSP problem of \autoref{sec:problem} on the road network. A structured query such as the one in \autoref{fig:parser-output} cannot be executed directly since its constraints vary from query to query and can be far more complex than in this example. Rather than encode this logic in the solver, we define a small set of functions through which any structured query becomes executable code, so a single fixed search procedure answers every query. This section first presents three \emph{core functions} that fully specify an SE-RCSP instance, and then \emph{optimization functions} that accelerate the search without changing its answer.

\subsection{Core Functions}
\label{sec:search}
The core solver receives the three functions below.

\smallskip
\noindent\textsc{Start}$(v_s) \rightarrow s_0$ returns the initial state at the source, with $f=0$ and $\mathbf{R}$ set to the values in the user's request, such as the starting battery.

\smallskip
\noindent\textsc{Expand}$(s,(u,v)) \rightarrow s'$ returns the state reached by traversing an edge: $f$ grows by the step cost $\delta$ and every resource is updated by $g$. If the new state breaks a step constraint in $C_{\text{step}}$, the move is not allowed and \textsc{Expand} returns $\bot$. Notice that the functions $\delta$ and $g$, and the step constraints are implicitly encoded within \textsc{Expand}.

\smallskip
\noindent\textsc{IsGoal}$(s)$ is true when the path reaching $s$ is \emph{feasible}, which by \autoref{eqn:feasible} is exactly when $s$ is at the destination $v_d$ and every trip constraint in $C_{\text{trip}}$ holds.

\begin{algorithm}[t]
\caption{Core SE-RCSP search}
\label{alg:core}
{
\footnotesize
\begin{algorithmic}[1]
\State $s_0 \gets \textsc{Start}(v_s)$ \Comment{State at the source} \label{alg:core:start}
\State OPEN $\gets \{s_0\}$ \Comment{Min-heap ordered by $f$} \label{alg:core:open}
\State $\mathit{best} \gets \bot$ \Comment{Best complete route so far} \label{alg:core:best}
\While{OPEN $\neq \emptyset$}
  \State $s \gets$ OPEN.pop() \Comment{Smallest $f$} \label{alg:core:pop}
  \If{\textsc{IsGoal}($s$)} \label{alg:core:isgoal}
    \If{$f$ is non-decreasing} \Comment{First goal is optimal} \label{alg:core:monotone}
      \State \Return \textsc{ReconstructPath}($s$) \Comment{Follow $\pi$ to the source} \label{alg:core:return}
    \ElsIf{$\mathit{best} = \bot$ \textbf{or} $s.f < \mathit{best}.f$} \label{alg:core:update}
      \State $\mathit{best} \gets s$ \Comment{Maximized: keep the best} \label{alg:core:keep}
    \EndIf
    \State \textbf{continue}
  \EndIf
  \ForAll{$(s.v,u) \in E$} \Comment{Outgoing edges of $s.v$} \label{alg:core:edges}
    \State $s' \gets \textsc{Expand}(s,(s.v,u))$ \label{alg:core:expand}
    \If{$s' = \bot$} \Comment{Step constraint violated}
      \State \textbf{continue}
    \EndIf
    \State $s'.\pi \gets s$; \ OPEN.insert($s'$) \label{alg:core:insert}
  \EndFor
\EndWhile
\State \Return \textsc{ReconstructPath}($\mathit{best}$) \textbf{if} $\mathit{best} \neq \bot$ \textbf{else} $\emptyset$ \label{alg:core:final}
\end{algorithmic}
}

\end{algorithm}

\autoref{alg:core} is the core search algorithm that uses these three functions and is inspired by Dijkstra's. It keeps OPEN, a priority queue of states not yet expanded, ordered by the accumulated objective $f$. Each iteration pops the cheapest state (line~\ref{alg:core:pop}) and tests it with \textsc{IsGoal} (line~\ref{alg:core:isgoal}). For a monotonically non-decreasing objective, e.g., distance or time, the first feasible path found is also optimal~\cite{irnich2005spprc}. For our cases, these are marked by a non-negative objective $f$.
The path is recovered by following the predecessor pointer $\pi$ back to the source (line~\ref{alg:core:monotone}). A maximization objective is negated, so $f$ is non-increasing and a state at the destination may be beaten by one that keeps accumulating negative cost. Therefore the search records the best complete route instead (line~\ref{alg:core:update}) and exhausts OPEN, which is bounded by other constraints, e.g., total trip time. If the goal is not reached, \textsc{Expand} is applied to every outgoing edge (line~\ref{alg:core:edges}), and the returned states are added to OPEN.

\subsection{Optimization}
\label{sec:efficiency}
The core search is correct but it could explore far more states than necessary. We introduce three mechanisms to reduce the work without changing the answer. \autoref{alg:full} is the optimized search with all optimization in place.


\smallskip
\noindent\textbf{Dominance.}
\textsc{Expand} discards a state that breaks a step constraint. Dominance further improves by discarding feasible states, when another state at the same node makes them redundant. For each node $v$, the solver keeps a frontier $\mathcal{F}[v]$ of the states that reached it, and calls \textsc{Dominates} to compare each new state against that frontier.

\begin{definition}[Dominance]
\label{def:dominance}
Let $s_a=(v,\mathbf{R}_a,f_a)$ and $s_b=(v,\mathbf{R}_b,f_b)$ be two states at the same node. $s_a$ \emph{dominates} $s_b$, written $s_a \preceq s_b$, if $f_a \le f_b$ and every resource in $\mathbf{R}_a$ is at least as good as the same resource in $\mathbf{R}_b$, with at least one of these comparisons strict. ``At least as good'' is defined per resource by the query, e.g., more battery left or less time elapsed; a resource with no such order must be equal in both states.
\end{definition}

\begin{algorithm}[t]
\caption{Optimized SE-RCSP search}
\label{alg:full}
{
\footnotesize
\begin{flushleft}
\textbf{Query-specific:} \textsc{Start}, \textsc{Expand}, \textsc{IsGoal}, \textsc{Dominates}; \\
\quad\quad\;\; \textsc{Expand} also sets each state's bound $\hat f$ (default $-\infty$) \\
\textbf{Output:} Optimal feasible path, or $\emptyset$ if none exists
\end{flushleft}
\begin{algorithmic}[1]
\State $s_0 \gets \textsc{Start}(v_s)$; \ OPEN $\gets \{s_0\}$
       \Comment{Ordered by $f+h(v)$} \label{alg:full:init}
\State $\mathcal{F}[s_0.v] \gets \{s_0\}$; \ $\mathit{best} \gets \bot$
       \Comment{Frontier, best so far} \label{alg:full:frontier}
\While{OPEN $\neq \emptyset$}
  \State $s \gets$ OPEN.pop()
  \If{\textsc{IsGoal}($s$)}
    \If{$f$ is non-decreasing} \Comment{First goal is optimal} \label{alg:full:monotone}
      \State \Return \textsc{ReconstructPath}($s$)
    \EndIf
    \State \textbf{continue} \Comment{Maximized: keep searching} \label{alg:full:incumbent}
  \EndIf
  \ForAll{$(s.v,u) \in E$}
    \State $s' \gets \textsc{Expand}(s,(s.v,u))$
         \Comment{$\bot$ if a step constraint breaks} \label{alg:full:prune}
    \If{$s' = \bot$} \textbf{continue}
    \EndIf
    \State $s'.\pi \gets s$
    \If{\textsc{IsGoal}($s'$)} \Comment{A complete route} \label{alg:full:inc}
      \If{$\mathit{best} = \bot$ \textbf{or} $s'.f < \mathit{best}.f$}
        \State $\mathit{best} \gets s'$ \Comment{New best}
      \EndIf
    \ElsIf{$\mathit{best} \neq \bot$ \textbf{and} $s'.\hat f \ge \mathit{best}.f$}
      \State \textbf{continue} \Comment{Cannot beat the best so far} \label{alg:full:bound}
    \EndIf
    \If{$\exists\, s_e \in \mathcal{F}[s'.v] : \textsc{Dominates}(s_e,s')$} \label{alg:full:dom1}
      \State \textbf{continue} \Comment{$s'$ is redundant} 
    \EndIf
    \State $\mathcal{F}[s'.v] \gets \{s_e \in \mathcal{F}[s'.v] : \neg\textsc{Dominates}(s',s_e)\} \cup \{s'\}$
           \label{alg:full:dom2}
    \State OPEN.insert($s'$)
  \EndFor
\EndWhile
\State \Return \textsc{ReconstructPath}($\mathit{best}$) \textbf{if} $\mathit{best} \neq \bot$ \textbf{else} $\emptyset$ \label{alg:full:ret}
\end{algorithmic}
}

\end{algorithm}

Dominance is safe when every continuation of $s_b$ to $v_d$ is also available from $s_a$ at no larger objective; discarding $s_b$ then cannot change the optimal objective. \autoref{def:dominance} is the usual Pareto test, and it is safe when each resource's order is preserved by $g$, $\delta$, and the constraints~\cite{irnich2005spprc}. Because those orders come from the query's own resources, \textsc{Dominates} is generated with the other functions rather than fixed in the solver (\autoref{fig:template-and-functions}). A new state dominated by one in the frontier is dropped, and any frontier state it dominates is removed (lines~\ref{alg:full:dom1}--\ref{alg:full:dom2}); states better in some components and worse in others are incomparable and both are kept.

\textsc{Dominates} only removes states, so a rule that prunes less than it could costs search time, not correctness. One that uses a wrong order, or ignores a resource, can discard the optimal route. Our experimental evaluation reports how far the returned routes are from the optimum.


\smallskip
\noindent\textbf{Bound pruning.}
Minimizing a non-increasing objective is expensive, since longer routes can yield better answers, e.g., by visiting more POIs. \emph{Bound pruning} uses the objective of the best valid answer found so far (line~\ref{alg:full:inc}) to prune routes that cannot beat it (line~\ref{alg:full:bound}).
This test is fixed in the solver. What it compares against is $\hat f$, an optimistic estimate of the best final objective any route through $s'$ can reach. The estimate depends on what the objective counts and on which resource limits it, so it is generated as part of \textsc{Expand}. For a collection query, the stops already collected plus the time left after reaching the destination, divided by the shortest stop duration, bound how many stops the route can end with. As with \textsc{Dominates}, a loose estimate costs only search time, while one that understates what is reachable can discard the optimum, and we do not verify it.

\smallskip
\noindent\textbf{Heuristic ordering.}
OPEN states (line~\ref{alg:full:init}) are ordered by the objective by default. This makes the search algorithm expand in all directions regardless of where the target is.
Adding a lower bound $h(v)$ on the cost from $v$ to $v_d$ and ordering by $f+h(v)$ expands states toward the destination first, as in A$^\star$. For a distance objective the straight-line distance to $v_d$ is such a bound, since no road is shorter than a straight line. Where the early return applies the bound must never overestimate, or the first accepted state may not be optimal; where the search drains OPEN instead, the order changes only how soon the best route is found. Changing the priority needs no new generated code, so we use $h=0$, plain Dijkstra, except in \autoref{sec:scalability}, which measures the gain.


\section{ORCStrator}
\label{sec:generator}
\begin{figure*}[!t]
\centering
\providecommand{\figfourbox}{\textwidth}
\providecommand{\figfourboxa}{0.40\textwidth}
\providecommand{\figfourboxb}{0.56\textwidth}
\providecommand{\figfourfont}{\footnotesize}
\providecommand{\figfourwrap}[1]{#1}

\figfourwrap{%
\begin{minipage}{\figfourbox}
\begin{minipage}[t]{\figfourboxa}
\vspace{0pt}
\centering
\textbf{(a)  Fixed abstract template with placeholder marker}\\[0.5em]

\figfourfont{
    \begin{minipage}[t]{\linewidth}
    \centering
    \begin{tabular}{@{}l@{}}
    \toprule
    \textbf{Template (fixed, with placeholders)} \\
    \midrule
    \texttt{class State:} \\
    \quad \texttt{node, prev, resources, objective} \\[2pt]
    \quad \texttt{start(origin, context):} \\
    \qquad \colorbox{gray!15}{\texttt{\_\_STATE\_START\_INIT\_BODY\_\_}} \\[4pt]
    \texttt{class Algorithm:} \\[2pt]
    \quad \texttt{expand(state):} \\
    \qquad \colorbox{gray!15}{\texttt{\_\_EXPAND\_BODY\_\_}} \\[4pt]
    \quad \texttt{dominates(a, b):} \\
    \qquad \colorbox{gray!15}{\texttt{\_\_DOMINATES\_BODY\_\_}} \\[4pt]
    \quad \texttt{is\_goal(state):} \\
    \qquad \colorbox{gray!15}{\texttt{\_\_IS\_GOAL\_BODY\_\_}} \\[4pt]
    \texttt{main():} \textit{--- fixed ---} \\
    \quad Load graph, build POI candidates \\
    \quad Run search with the four methods \\
    \quad Write output \\
    \bottomrule
    \end{tabular}
    \end{minipage}%
    \hfill
}


\end{minipage}
\hfill
\begin{minipage}[t]{\figfourboxb}
\vspace{0pt}
\centering
\textbf{(b) Generated functions filled by the LLM}\\[0.5em]
\figfourfont
\input{algorithms/four-functions}
\end{minipage}
\end{minipage}}

\caption{Template and generated functions for the running example.}
\label{fig:template-and-functions}

\end{figure*}

\label{sec:template}

The ORCStrator bridges the Spec Builder and the SE-RCSP Solver, turning the structured query of \autoref{sec:input-parser} into the four functions that the solver (\autoref{alg:full}) executes, so one fixed solver answers any query. An expert could hand-code these functions for a single query type, but this would be impractical given the virtually unlimited combinations of constraints and objectives. The ORCStrator instead generates them with an LLM, so new combinations are supported as soon as a user asks for them. Only these four functions are generated and the solver, graph loading, and output logic stay fixed and outside the prompt, leaving the model little room to go wrong.


\smallskip
\subsection{Prompt Assembly}
The prompt tells the model what this query needs and what it may use, without exposing the solver. It has four parts:

\smallskip \noindent \textbf{Input.} The Spec Builder output of \autoref{sec:input-parser} and the original NL query, with the POI and road schema, so the code refers only to categories, attributes, and road classes in the data.

\smallskip \noindent \textbf{Definitions.} What each function must return and the \textsc{State} fields it may touch. The objective and the constraints are already fixed by the Spec Builder; the model writes the code that enforces them, with a maximized objective stored negated so the same minimizing solver applies.

\smallskip \noindent \textbf{Helper functions.} The parts of the solver the generated code may call rather than rewrite, given by name and purpose: reading an edge weight, an optimistic travel time to the destination, and \textsc{BoundPrune} (\autoref{alg:full}, lines~\ref{alg:full:inc}--\ref{alg:full:bound}) for a maximized objective.

\smallskip \noindent \textbf{Rules and output format.} The four signatures with their return contracts, for instance that \textsc{Expand} yields \textsc{State} objects and \textsc{Dominates} returns a boolean, and a JSON response with one key per function body.

The code of the solver is never passed in this step. The model is only told in words what a state holds and what each function must return. This level of abstraction allows us to improve the solver algorithms in the future without having to change the ORCStrator.

\smallskip
\subsection{Filling the Template}
The LLM returns the four bodies, which are substituted into the template's placeholders. \autoref{fig:template-and-functions} shows the result for our running example, in pseudo code; the system generates Python. \textsc{Start} declares what this query tracks, the charge left and how many required stops have been visited (\texttt{battery} and \texttt{poi\_index} in the figure). \textsc{Expand} updates both along each edge, draining the battery, restoring it at a charging station, and advancing the stop counter at a matching neighbor. \textsc{Dominates} compares two states on the objective and both resources, and \textsc{IsGoal} accepts only at the destination with every stop visited. A different query changes only what these four carry: a time budget adds elapsed time as a third resource, and nothing else changes.

\subsection{Matching Stops}
\label{sec:filtering}
Most filters are answered by the POI table directly: in the running example, \emph{a Thai restaurant} is just the \emph{cuisine} attribute. \emph{A park with a playground} has no such column, so CRAFT matches the phrase against the POI's text description instead, in two steps. POI descriptions are embedded offline. At query time CRAFT embeds the requested phrase and compares the two vectors by cosine similarity: POIs below a threshold $\theta$ are dropped as clear non-matches, and a local LLM reads the rest and answers yes or no; only the LLM can accept a POI, since similarity alone is too unreliable to decide. Each POI is checked the first time \textsc{Expand} reaches it, and the answer is stored and reused if the search reaches it again. Section~\ref{sec:experiments} reports how $\theta$ is chosen and what the check costs.


\begin{figure}[!tb]
\centering
\footnotesize
\begin{tabular}{@{}ll@{}}
\toprule
\textbf{Field} & \textbf{Value} \\
\midrule
query\_id    & example\_query\_1 \\
route        & [v1, v4, v5, v8, v7, v9, v10] \\
poi\_stops   & [\{type: charger, node: v4\}, \\
             & \;\{type: park, node: v7\}, \\
             & \;\{type: restaurant, node: v9\}] \\
objective(normalized to meters)    & 27358.8 \\
weight\_key  & distance \\
run\_time\_ms & 12.34 \\
\bottomrule
\end{tabular}

\caption{Search output for the example in~\autoref{fig:toy_example}.}
\label{fig:exec-output}
\end{figure}

\section{Benchmark Design}
\label{sec:benchmark}

Existing benchmarks cover only part of CRAFT's query space. The closest, LLMAP~\cite{yuan2025llmap}, uses real map data but only one query type, that is, multi-stop trips over five fixed POI types, with no battery, ordering, distance, or edge constraints. Others use synthetic grid maps~\cite{wang2025routellm} or plan multi-day travel without a road network~\cite{xie2024travelplanner, shao2026chinatravel}. Therefore, we build our own benchmark, including LLMAP's queries re-grounded on our network for direct comparison.


\noindent\textbf{Data Preparation}
We extract road networks from OpenStreetMap using OSMnx~\cite{boeing2017osmnx} at two scales: a Southern California (SoCal) network (248,106 nodes, 635,056 edges, 627,220 POIs), and a statewide California (CA) network (1,186,063 nodes, 2,957,324 edges, 2,258,287 POIs) for scalability. Edges carry length, travel time, and road class; POIs carry category labels and structured attributes such as cuisine, wheelchair accessibility, and opening hours. We keep only the largest strongly connected component for full routability, and each POI is snapped to its nearest network node.

To support the description check (Section~\ref{sec:filtering}), we enrich each POI with a text description, taken from its OSM description tag, else from linked Wikipedia or Wikidata entries, else generated by an LLM from its OSM attributes. This yields 87,316 described POIs in SoCal, pre-embedded with OpenAI's text-embedding-3-small~\cite{openembedding3} for query-time matching.

\smallskip\noindent\textbf{Query Sets}
\autoref{tab:templates} lists the nine query sets with an example of each, and \autoref{tab:queires} summarizes their objectives and constraints. The first five cover the problem families of \autoref{sec:related_Work}: (Q1-Q4) are route-planning templates with mandatory requirements, built from battery tracking~\cite{baum2019chargingstops, storandt2012cruising}, unordered stops~\cite{li2005trip}, ordered stops~\cite{sharifzadeh2008osr, zhu2022rcosr}, and their combination under per-leg limits; and (Q5) covers POI Collection orienteering problem~\cite{vansteenwegen2011orienteering, bolzoni2014} that fits as many stops as possible into a time budget. 
Three \emph{specialized templates} extend it: (Q6) highway avoidance, which constrains edge attributes rather than POIs~\cite{delling2017crp, zhao2024semantic}; (Q7) attribute and description filters, which ask for stops that plain categories cannot express; and (Q8) long trips, which reuse EV Routing on the statewide network. Finally, (Q9) builds on LLMAP~\cite{yuan2025llmap} using our SoCal dataset since its road network is not available.

We instantiate 200 queries per template (Q1-Q8) and 1,000 for (Q9). We sample a source and destination and place the required stops at real POIs so the query is feasible. In some EV queries the battery is set below what the direct trip needs, so a valid route must pass a charging station.

\begin{table}[t]
\centering
\caption{Benchmark query sets. Main and specialized sets are introduced in this paper; the external set (LLMAP) is re-grounded onto our road network. RP = route planning, IP = itinerary planning.}\label{tab:templates}

\footnotesize
\setlength{\tabcolsep}{3pt}
\begin{tabular}{@{}llrl@{}}
\toprule
& \textbf{Query set} & \textbf{Count} & \textbf{Family} \\
\midrule
\multicolumn{4}{@{}l}{\emph{Main}} \\
Q1 & EV Routing~\cite{baum2019chargingstops, storandt2012cruising} & 200 & RP \\
\multicolumn{4}{@{}p{0.95\columnwidth}}{\scriptsize\itshape\quad ``What's the shortest route from $A$ to $B$? I'm at 12\% of a 250-mile battery.''} \\[2pt]
Q2 & EV + Ordered~\cite{baum2019chargingstops, sharifzadeh2008osr} & 200 & RP \\
\multicolumn{4}{@{}p{0.95\columnwidth}}{\scriptsize\itshape\quad ``I need the shortest route from $A$ to $B$ with 20\% charge: a pharmacy first and then a bank, with no leg over 5 miles.''} \\[2pt]
Q3 & Unordered POI~\cite{li2005trip} & 200 & RP \\
\multicolumn{4}{@{}p{0.95\columnwidth}}{\scriptsize\itshape\quad ``What's the fastest way from $A$ to $B$? I want a supermarket and a pharmacy in any order, with each leg within 11 miles.''} \\[2pt]
Q4 & Ordered POI~\cite{sharifzadeh2008osr} & 200 & RP \\
\multicolumn{4}{@{}p{0.95\columnwidth}}{\scriptsize\itshape\quad ``What's the shortest route from $A$ to $B$, stopping at a pharmacy within 7 miles, then a library about 2 miles after that?''} \\[2pt]
Q5 & POI Collection~\cite{vansteenwegen2011orienteering, bolzoni2014} & 200 & IP \\
\multicolumn{4}{@{}p{0.95\columnwidth}}{\scriptsize\itshape\quad ``From $A$ to $B$ in 2 hours, visiting as many places as possible: up to 2 pharmacies, 1 library, and 1 bank.''} \\
\midrule
\multicolumn{4}{@{}l}{\emph{Specialized}} \\
Q6 & Unordered + Hwy.~\cite{delling2017crp, zhao2024semantic} & 200 & RP \\
\multicolumn{4}{@{}p{0.95\columnwidth}}{\scriptsize\itshape\quad ``What's the fastest route from $A$ to $B$ via a pharmacy, avoiding highways?''} \\[2pt]
Q7 & Ordered POI + Desc.~\cite{sharifzadeh2008osr} & 200 & RP \\
\multicolumn{4}{@{}p{0.95\columnwidth}}{\scriptsize\itshape\quad ``What's the shortest way from $A$ to $B$ with two stops: first a Thai restaurant within 5 miles, then an art museum about 5 miles later?''} \\[2pt]
Q8 & Long Trips~\cite{baum2019chargingstops, storandt2012cruising} & 200 & RP \\
\multicolumn{4}{@{}p{0.95\columnwidth}}{\scriptsize\itshape\quad ``What's the shortest drive from $A$ to $B$? My EV is at 30\% of a 300-mile range.''} \\
\midrule
\multicolumn{4}{@{}l}{\emph{External}} \\
Q9 & LLMAP~\cite{yuan2025llmap} & 1{,}000 & RP \\
\multicolumn{4}{@{}p{0.95\columnwidth}}{\scriptsize\itshape\quad ``On my way from A to B, let's visit a bank, a library, and a shopping mall.''} \\
\midrule
& \textbf{Total} & \textbf{2{,}600} & \\
\bottomrule
\end{tabular}
\end{table}

\begin{table}[t]
\centering
\caption{Objective and constraints of each query set. Objectives are minimized except Q5, which maximizes number of POIs.}\label{tab:queires}
\footnotesize
\setlength{\tabcolsep}{2pt}
\begin{tabular}{@{}lcccccccc@{}}
\toprule
& \textbf{Objective} & \textbf{Battery} & \textbf{Stops} & \textbf{Order} & \textbf{Distance} & \textbf{Highway} & \textbf{Time} & \textbf{Filters} \\
\midrule
Q1 & distance & \checkmark & & & & & & \\
Q2 & distance & \checkmark & \checkmark & \checkmark & \checkmark & & & \\
Q3 & time & & \checkmark & & \checkmark & & & \\
Q4 & distance & & \checkmark & \checkmark & \checkmark & & & \\
Q5 & POIs & & \checkmark & & & & \checkmark & \\
Q6 & time & & \checkmark & & & \checkmark & & \\
Q7 & distance & & \checkmark & \checkmark & \checkmark & & & \checkmark \\
Q8 & distance & \checkmark & & & & & & \\
Q9 & distance & & \checkmark & \checkmark & & & \checkmark & \checkmark \\
\bottomrule
\end{tabular}
\end{table}

\section{Experiments}
\label{sec:experiments}
\subsection{Experimental Setup}
\label{sec:setup}
We evaluate CRAFT on the 2,600 queries of~\autoref{tab:templates} with three LLM backends, Gemini 3.1 Flash Lite~\cite{flashlite} and GPT-4.1~\cite{gpt4} via their commercial APIs, and the open-weight gpt-oss-20b~\cite{gptoss} on a local NVIDIA RTX 5090; one model serves all LLM calls within a run. We also tried a newer model, GPT-5.5~\cite{gpt55} but it performed close to GPT 4.1 for our queries so we used the older and cheaper model.

\subsubsection{Baselines}
\label{sec:baselines}
To the best of our knowledge, no existing system answers this range of constraints end to end from natural language. We therefore compare against the two closest: both take the query as free text and return stops on real map data instead of requiring a structured query.

\smallskip
\noindent\textbf{I. LLM-Only Baseline.}
The simplest baseline prompts an LLM with the query to return a verifiable sequence of POI stops, each with a name, address, and type. To account for hallucination, we use Google Geocoding and Places APIs~\cite{google-geocoding} and take the nearest same-type POI within 200 meters, breaking ties by exact name. Even if the POI is not in our dataset, we consider it correct for evaluation and snap it to the nearest graph node. Then, we connect these stops with shortest path to reconstruct the route for evaluation.

\smallskip
\noindent\textbf{II. LLMAP Baseline.}
LLMAP~\cite{yuan2025llmap} uses LLM to extract POI types and ordering, then enumerates type-group permutations over a candidate-POI graph and runs Dijkstra on each until one is feasible. LLMAP supports only five hardcoded POI types with fixed visit durations and travel speed~\cite{elmahallawy2026adaptive}.

\subsubsection{Evaluation Metrics}
\label{sec:metrics}
\noindent\textbf{Correctness.} Every returned route is verified by a hand-written checker that replays it on the graph and tests each constraint of its query. A route passes only if all constraints hold. We included 50 queries with no valid answer; returning no route for them is considered correct (\emph{True Infeasible}, T-Inf). Failures are reported in three categories: \emph{Error}, for code errors, crashes, timeouts, and hallucinated stops; \emph{False Infeasible} (F-Inf), for returning no route when one exists; and \emph{Violation} (Viol), for a route that breaks a constraint, including any route returned for an infeasible query.

\smallskip
\noindent\textbf{Objective quality.} For each query that at least one system answers correctly, we rank the systems by objective, and score each system by how often it ranks first. Only CRAFT returns a full route, so we connect the baselines' stops with shortest paths and use that cost as their objective.

\smallskip
\noindent\textbf{Optimality.} We compare CRAFT's correct routes against optimal answers that we compute using a brute-force algorithm (Section~\ref{sec:results}).

\smallskip
\noindent\textbf{Scalability.} We report median wall-clock search time and number of expanded states over the queries that finished, excluding graph loading, which is the same for all systems, and LLM API latency, which is not comparable across hardware.


\subsection{End-to-end Experiments}
\label{sec:results}

\begin{table}[t]
\centering
\caption{Correct routes for the main Query Set and failure breakdown}
\label{tab:results}
\scriptsize
\setlength{\tabcolsep}{2pt}
\begin{tabular}{@{}ll | rrrrr r r | rrr @{}}
\toprule
& & \multicolumn{7}{c|}{\textbf{Correct}} & \multicolumn{3}{c}{\textbf{Incorrect}} \\
\cmidrule(lr){3-9} \cmidrule(l){10-12}
\textbf{Model} & \textbf{System}
 & \rotatebox{70}{EV Routing} & \rotatebox{70}{EV+Ordered} & \rotatebox{70}{Unord.\ POI}
 & \rotatebox{70}{Ordered POI} & \rotatebox{70}{POI Collect.} & \rotatebox{70}{T-Inf}
 & \rotatebox{70}{\textbf{Total}} & Error & F-Inf & Viol \\
\midrule
\multirow{3}{*}{Flash Lite}
& LLM   & 89 & 20 & 32 & 8 & 51 & 0 & 200 & 616 & 1 & 183 \\
& LLMAP & 92 & 79 & 170 & 52 & 27 & 0 & 420 & 20 & 153 & 407 \\
& CRAFT & \textbf{200} & \textbf{185} & \textbf{181} & \textbf{124} & \textbf{172} & \textbf{49} & \textbf{911} & 38 & 20 & 31 \\
\midrule
\multirow{3}{*}{GPT-4.1}
& LLM   & 98 & 19 & 13 & 11 & 123 & 1 & 265 & 542 & 0 & 193 \\
& LLMAP & 92 & 80 & 174 & 48 & 27 & 0 & 421 & 2 & 171 & 406 \\
& CRAFT & \textbf{194} & \textbf{192} & \textbf{178} & \textbf{138} & \textbf{184} & \textbf{48} & \textbf{934} & 25 & 26 & 15 \\
\midrule
\multirow{3}{*}{gpt-oss}
& LLM   & 83 & 0 & 1 & 0 & 78 & 38 & 200 & 309 & 415 & 76 \\
& LLMAP & 92 & 78 & \textbf{166} & 44 & 28 & 0 & 408 & 33 & 139 & 420 \\
& CRAFT & \textbf{178} & \textbf{174} & 153 & \textbf{127} & \textbf{143} & \textbf{49} & \textbf{824} & 118 & 34 & 24 \\
\bottomrule
\end{tabular}

\end{table}

\smallskip
\noindent\textbf{Correctness.}
\autoref{tab:results} reports the correctness for the 1,000 queries in the main query set (Q1-Q5) and breaks down the incorrect answers by type.

\smallskip
\noindent\emph{CRAFT} leads the correctness results across all templates with an overall rate of up-to 93\%. Interestingly, even with the open-weight model~\cite{gptoss}, it scores 82\% correctness rate. In addition, CRAFT can accurately recognize almost all 50 invalid queries (T-Inf) because it runs every query on the actual road network. Failures are dominated by code errors including code that does not compile or does not finish on time and the open-weight model producing significantly more code errors. Other types of errors stay around 3\% for constraint violations and failing to find any answer even though one exists.


\smallskip
\noindent\emph{The LLM-only baseline} is correct on 20--27\% of queries. Its failures are overwhelmingly errors due to hallucinations, e.g., non-existent businesses or a fabricated address.
The backends differ in temperament rather than in accuracy. Flash Lite and GPT-4.1 are over-confident, almost never declining a query even when it has no answer. GPT-oss is the opposite, declining 453 times, but it is wrong on 415 of them, so caution does not help it either.

\smallskip

\noindent\emph{LLMAP} is correct on about 41\% of queries, peaking on Unordered POI, the family it was built for. Because only its parsing step uses an LLM, it is the most stable across backends. Since it was not designed for EV routing, it answers exactly the 92 queries that need no charging stop. The errors are mainly due to parsing errors since LLM does not generate any code. LLMAP never reports a true infeasible because the infeasibility is due to per-leg distance limit which LLM does not capture. On the other hand, it returns false infeasible for many POI Collection queries since it does not know how to process them.

\noindent\textbf{Objective Quality.}
\autoref{tab:ranking} compares route quality. A query is credited to whichever systems returned a correct route with the best objective, the lowest cost for Q1--Q4 or the most visited POIs for Q5. \emph{Failed} indicates the queries that all systems failed to answer. Systems that reach the same objective are all credited, which happens on about a fifth of the answered queries, so a row can sum to more than \#Q. CRAFT ranks first on 77\%--88\% of the queries depending on the model, and leads every template with every model. The baselines win where their design is strongest: LLMAP on Unordered POI (Q3), the family it was built for, and the LLM-only baseline on POI Collection (Q5), where the large search space makes the template hardest for everyone else.

\begin{table}[t]
\centering
\caption{Quality Experiments. Report the number of queries that each system answered correctly and with the best objective. Failed = all systems failed.}
\label{tab:ranking}

\footnotesize
\setlength{\tabcolsep}{3pt}
\begin{tabular}{@{}l l r r r r r@{}}
\toprule
\textbf{Model} & \textbf{Template} & \textbf{\#Q} & \textbf{LLM} & \textbf{LLMAP} & \textbf{CRAFT} & \textbf{Failed} \\
\midrule
\multirow{6}{*}{Flash Lite}
& EV Routing       & 200 & 89 & 92 & \textbf{200} &   0 \\
& EV + Ordered     & 200 & 1 & 28 & \textbf{185} &  14 \\
& Unordered POI    & 200 & 1 & 38 & \textbf{179} &   4 \\
& Ordered POI      & 200 & 1 & 32 & \textbf{124} &  68 \\
& POI Collection   & 200 & 27 & 4 & \textbf{168} &  18 \\
\cmidrule{2-7}
& \textbf{Total} & 1000 & 119 & 194 & \textbf{856} & 104 \\
\midrule
\multirow{6}{*}{GPT-4.1}
& EV Routing       & 200 & 96 & 92 & \textbf{194} &   4 \\
& EV + Ordered     & 200 & 0 & 28 & \textbf{192} &   7 \\
& Unordered POI    & 200 & 0 & 47 & \textbf{175} &   3 \\
& Ordered POI      & 200 & 4 & 28 & \textbf{138} &  55 \\
& POI Collection   & 200 & 50 & 0 & \textbf{181} &  10 \\
\cmidrule{2-7}
& \textbf{Total} & 1000 & 150 & 195 & \textbf{880} & 79 \\
\midrule
\multirow{6}{*}{gpt-oss-20b}
& EV Routing       & 200 & 83 & 92 & \textbf{178} &  14 \\
& EV + Ordered     & 200 & 0 & 34 & \textbf{174} &  19 \\
& Unordered POI    & 200 & 1 & 53 & \textbf{153} &  11 \\
& Ordered POI      & 200 & 0 & 26 & \textbf{127} &  68 \\
& POI Collection   & 200 & 28 & 8 & \textbf{141} &  35 \\
\cmidrule{2-7}
& \textbf{Total} & 1000 & 112 & 213 & \textbf{773} & 147 \\
\bottomrule
\end{tabular}

\end{table}

\begin{figure}[t]
\centering
\input{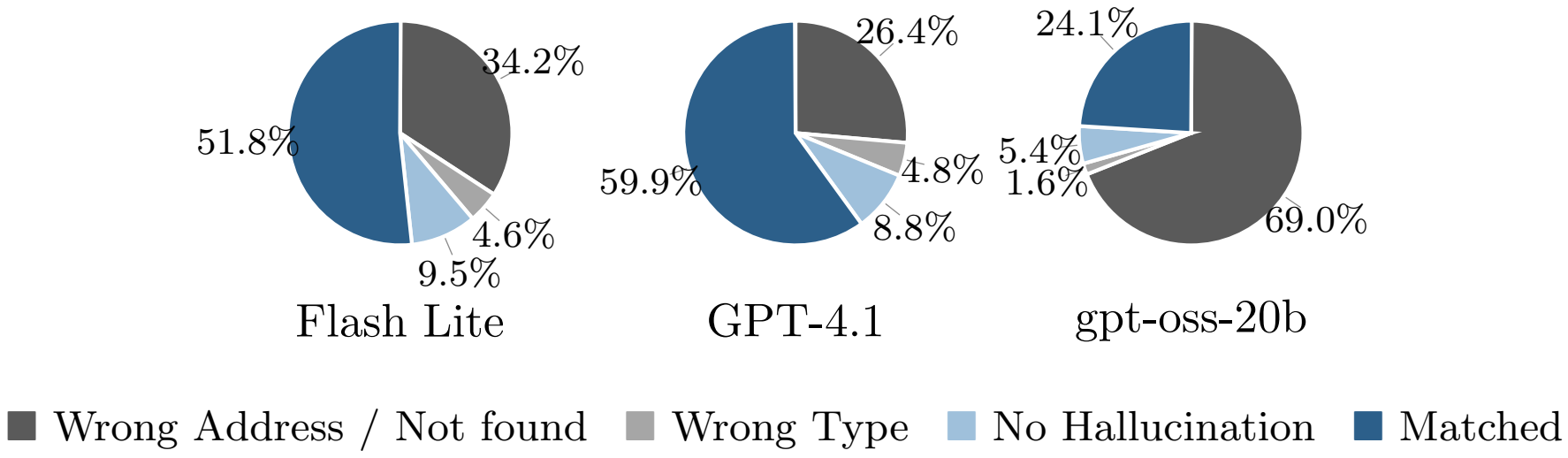}
\caption{Per-stop classification of the stops proposed by the LLM-only baseline.}
\label{fig:hallucination}
\end{figure}

\noindent\textbf{Failures Breakdown.}
LLM-only relies on the internal knowledge of the model so we have to verify the correctness of the POIs in the result. Since OSM dataset is incomplete, we use Google Places API to verify whether a business with the suggested name exists, whether it sits at the claimed address, and whether it matches the requested type.
\autoref{fig:hallucination} breaks down that check over every stop the baseline proposed. The two commercial models place most of their stops correctly: 51.8\% and 59.9\% resolve to a real POI in our data, and these are the stops on which we can compare the baseline fairly with the other systems. The open-weight model is far weaker, matching only 24.1\%, with 69\% of its stops at an address Google cannot confirm or far from where it claimed. Claiming the wrong type of place is rare everywhere, at most 4.8\%. The remaining 5--10\% are \emph{No Hallucination}: real, correctly located places that are simply missing from our OSM extract, which we snap to the nearest node so the route can still be scored.

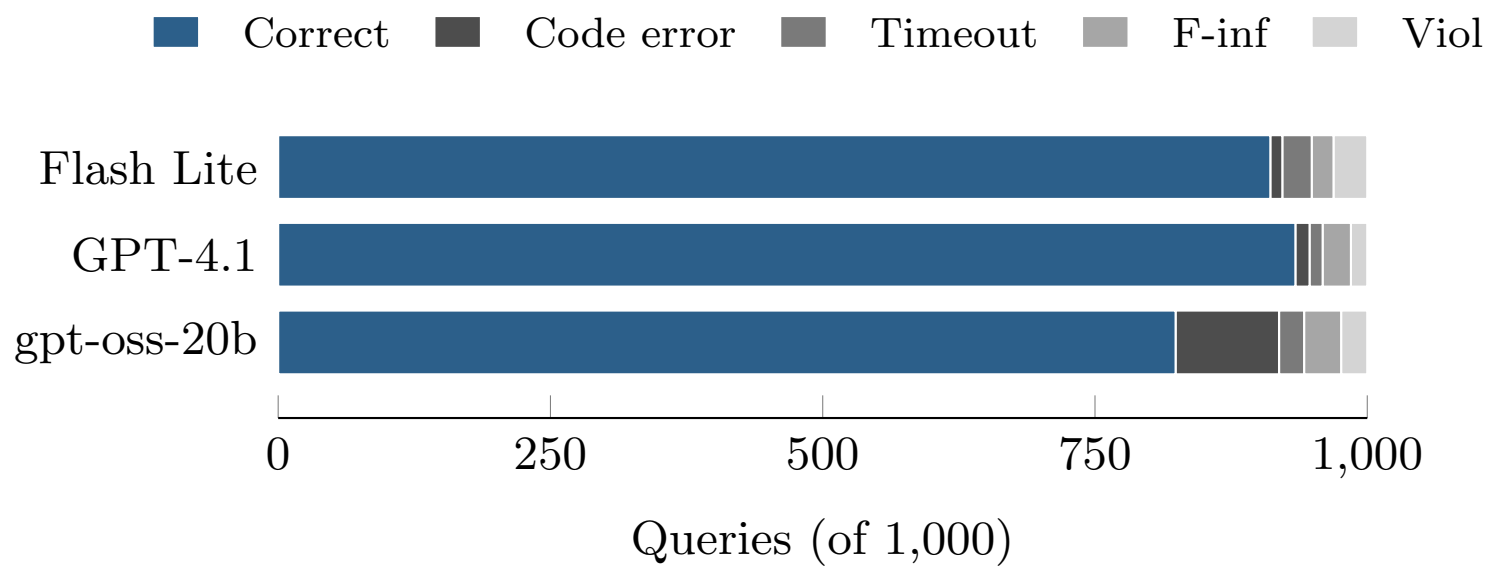
\begin{figure}[t]
\centering
\definecolor{fCorrect}{HTML}{2C5F8A} 
\definecolor{fCode}{HTML}{4D4D4D}    
\definecolor{fTimeout}{HTML}{7A7A7A} 
\definecolor{fMiss}{HTML}{A6A6A6}    
\definecolor{fViol}{HTML}{D4D4D4}    
\begin{tikzpicture}
\begin{axis}[
    xbar stacked, width=0.74\craftfigbox, height=3.4cm,
    xmin=0, xmax=1000, bar width=12pt,
    enlarge y limits={abs=0.5cm}, y=0.58cm,
    axis x line*=bottom,
    y axis line style={draw=none}, ytick style={draw=none},
    xtick={0,250,500,750,1000}, tick label style={font=\footnotesize},
    xlabel={\footnotesize Queries (of 1{,}000)},
    symbolic y coords={gpt-oss-20b,GPT-4.1,Flash Lite}, ytick=data,
    yticklabel style={font=\footnotesize},
    legend style={at={(0.5,1.04)}, anchor=south, legend columns=-1,
        draw=none, font=\scriptsize, column sep=6pt},
    every axis plot/.append style={draw=white, line width=0.4pt},
]
\addplot[fill=fCorrect] coordinates {(911,Flash Lite) (934,GPT-4.1) (824,gpt-oss-20b)};
\addplot[fill=fCode]    coordinates {(11,Flash Lite) (13,GPT-4.1) (95,gpt-oss-20b)};
\addplot[fill=fTimeout] coordinates {(27,Flash Lite) (12,GPT-4.1) (23,gpt-oss-20b)};
\addplot[fill=fMiss]    coordinates {(20,Flash Lite) (26,GPT-4.1) (34,gpt-oss-20b)};
\addplot[fill=fViol]    coordinates {(31,Flash Lite) (15,GPT-4.1) (24,gpt-oss-20b)};
\legend{Correct, Code error, Timeout, F-inf, Viol}
\end{axis}
\end{tikzpicture}
\caption{CRAFT outcome breakdown per LLM model (1{,}000 queries)}
\label{fig:craft-failures}
\end{figure}
\autoref{fig:craft-failures} dissects CRAFT's incorrect results. The error mix shifts with the model: with Flash Lite most are timeouts, 27 of 38; with GPT-4.1 timeouts and code errors are even, 12 against 13; and with gpt-oss-20b code errors dominate, 95 of 118, most of them code that never runs. False infeasibility (F-Inf), where CRAFT returns no route although the exact solver found one, stays at 3.5\% or less, as do violations.

\subsection{LLMAP Queries}
Table~\ref{tab:llmapset} reports results on the 1,000 re-grounded LLMAP queries (Q9), evaluated with the same pipeline as the main benchmark. On its own query distribution, LLMAP is strong, answering about three quarters of the queries correctly with every model, yet CRAFT still produces at least as many correct routes (761, 924, and 764) and wins the objective comparison decisively, ranking first on at least 695 queries against at most 279 for LLMAP. The gap between matching LLMAP on correctness and dominating it on first-rank reflects route quality: LLMAP returns the first feasible itinerary rather than the best one. CRAFT's violations remain minimal (11, 1, and 4), and the LLM-only baseline collapses on this set.
\begin{table}[t]
\centering
\caption{Results of LLMAP-inspired benchmark of 1{,}000 queries~\cite{yuan2025llmap}. First-rank credits every system tied on the best objective, as in \autoref{tab:ranking}.}
\label{tab:llmapset}

\footnotesize
\setlength{\tabcolsep}{3pt}
\begin{tabular}{@{}l l r r r r r@{}}
\toprule
\textbf{Model} & \textbf{System} & \textbf{Correct} & \textbf{Fail} & \textbf{Infeas} & \textbf{Viol} & \textbf{First-rank} \\
\midrule
\multirow{3}{*}{Flash Lite}
 & LLM   & 134 & 625 & 10 & 231 & 19 \\
 & LLMAP & 757 & 0 & 24 & 219 & 279 \\
 & CRAFT & \textbf{761} & 48 & 180 & 11 & \textbf{695} \\
\midrule
\multirow{3}{*}{GPT-4.1}
 & LLM   & 150 & 580 & 78 & 192 & 11 \\
 & LLMAP & 775 & 0 & 3 & 222 & 188 \\
 & CRAFT & \textbf{924} & 24 & 51 & 1 & \textbf{846} \\
\midrule
\multirow{3}{*}{gpt-oss-20b}
 & LLM   & 31 & 241 & 528 & 200 & 0 \\
 & LLMAP & 754 & \textbf{0} & 30 & 216 & 278 \\
 & CRAFT & \textbf{764} & 159 & 73 & 4 & \textbf{701} \\
\bottomrule
\end{tabular}

\end{table}

\noindent\textbf{Optimality}
To measure how close CRAFT's routes are to the best possible ones, we built a brute-force exact solver for each of the five templates (Q1-Q5). The solver tries every permutation of stops satisfying the query's constraints, using a precomputed matrix of shortest-path distances and travel times between candidate stops, and returns the one with the best objective. We then compare each correct CRAFT route against this optimum and report the mean absolute percentage error (MAPE) of the objective, $\lvert f - f^{*} \rvert / f^{*}$ averaged over correct routes, where $f$ is the route's objective and $f^{*}$ the optimum. Table~\ref{tab:optimality} shows the results per template and model.
The error is very low on the route-planning templates, at most 2.21\% for any model. POI Collection is the exception, since its much larger search space makes the optimum harder to reach: the mean error rises to 6.98\% with Flash Lite and 14.98\% with gpt-oss-20b, while GPT-4.1 stays at 1.40\%.

\begin{table}[t]
\centering
\caption{Optimality against the exact solver: MAPE of correct routes' objectives relative to the proven optimum, per template and model.}
\label{tab:optimality}

\footnotesize
\begin{tabular}{l ccc}
\toprule
Template & Gemini Flash Lite & GPT-4.1 & gpt-oss-20b \\
\midrule
EV Routing     & 0.62\% & 0.24\% & 0.27\%  \\
EV + Ordered   & 0.77\% & 0.61\% & 1.40\%  \\
Ordered POI    & 0.10\% & 0.10\% & 0.33\%  \\
Unordered POI  & 2.21\% & 1.34\% & 1.39\%  \\
POI Collection & 6.98\% & 1.40\% & 14.98\% \\
\bottomrule
\end{tabular}

\end{table}

\subsection{Specialized Queries}
\label{sec:specialized}
This part studies CRAFT with two specialized queries, Q6 and Q7. Since neither baseline supports these queries, it would be difficult to run a fair evaluation on them.

\noindent\textbf{Highway Avoidance (Q6)}
The queries in Q6 extends Q3 by adding a highway avoidance constraint, which tests constraints on attributes on edges.Since some routes are impossible to navigate without highways in our datasets, a fixed threshold would be unfair, so we score the constraint by comparison. For each answer we take the stops CRAFT visited and compute the shortest path through them with no highway preference at all, which is the route a system that ignored the constraint would return. We count the answer correct when CRAFT's route uses less highway than that baseline, so the test asks whether the generated code acted on the constraint, and it stays fair across trips that differ in how much highway they can avoid.

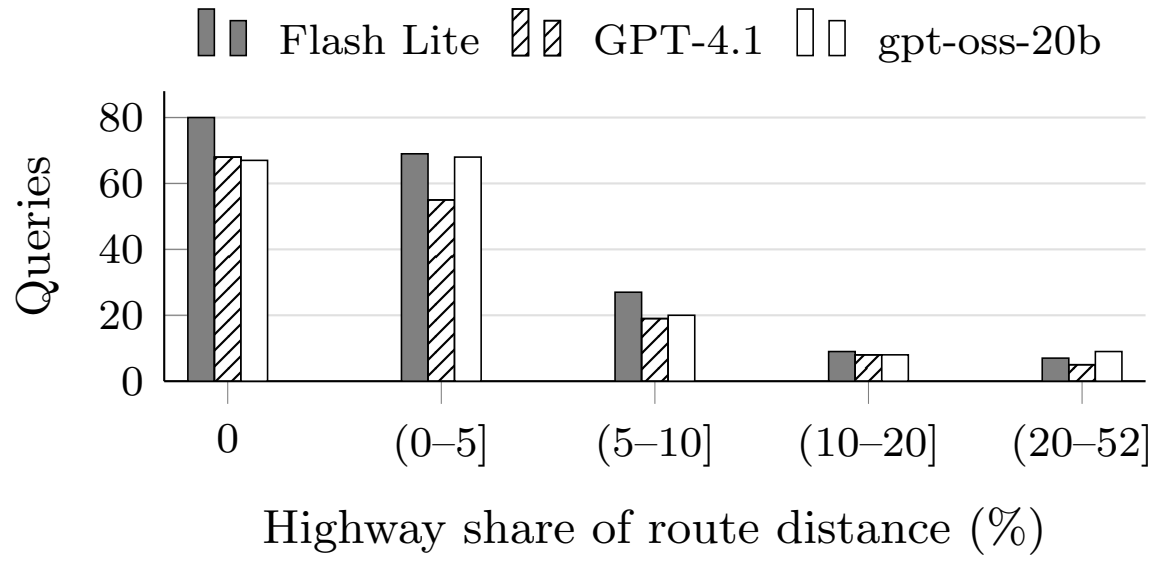
\begin{figure}[t]
\centering
\begin{minipage}{\craftfigbox}
\begin{minipage}[b]{0.30\craftfigbox}
\centering
\footnotesize
\setlength{\tabcolsep}{4pt}
\begin{tabular}{@{}l r@{}}
\toprule
\textbf{Model} & \textbf{Correct} \\
\midrule
Flash Lite  & \textbf{192} \\
GPT-4.1     & 155 \\
gpt-oss-20b & 172 \\
\bottomrule
\end{tabular}
\vspace{0.9cm}
\end{minipage}%
\hfill
\begin{minipage}[b]{0.68\craftfigbox}
\centering
\begin{tikzpicture}
\begin{axis}[
    ybar, width=\linewidth, height=3.5cm,
    bar width=5pt, bar shift auto=0pt,
    enlarge x limits={abs=0.42cm},
    xtick pos=bottom, ytick pos=left,
    ymin=0, ymax=88, ytick={0,20,40,60,80},
    ylabel={\footnotesize Queries},
    xlabel={\footnotesize Highway share of route distance (\%)},
    tick label style={font=\scriptsize}, label style={font=\footnotesize},
    ymajorgrids, grid style={draw=black!12},
    axis x line*=bottom, axis y line*=left,
    symbolic x coords={0,(0--5],(5--10],(10--20],(20--52]},
    xtick=data, xticklabel style={font=\scriptsize},
    legend style={at={(0.5,1.02)}, anchor=south, legend columns=-1,
        draw=none, font=\scriptsize, column sep=4pt},
]
\addplot[fill=gray, draw=black, line width=0.3pt] coordinates {
  (0,80) ((0--5],69) ((5--10],27) ((10--20],9) ((20--52],7)};
\addplot[fill=white, postaction={pattern=north east lines, pattern color=black},
         draw=black, line width=0.3pt] coordinates {
  (0,68) ((0--5],55) ((5--10],19) ((10--20],8) ((20--52],5)};
\addplot[fill=white, draw=black, line width=0.3pt] coordinates {
  (0,67) ((0--5],68) ((5--10],20) ((10--20],8) ((20--52],9)};
\legend{Flash Lite, GPT-4.1, gpt-oss-20b}
\end{axis}
\end{tikzpicture}
\end{minipage}
\end{minipage}
\caption{Highway-avoidance template: correct routes per backbone (left) and the distribution of highway share of route distance (right). The leftmost bin is routes using no highway at all; the highest share observed is 52\%.}
\label{fig:hwy}
\end{figure}

Figure~\ref{fig:hwy} shows the number of correct routes out of the 200 queries. CRAFT performs well on all models with Flash Lite giving the best answer. We also plot a histogram of the fraction of the route (by distance) on highway. About 40\% of the routes use no highway at all, and nearly 80\% keep their highway share below 5\%.

\noindent\textbf{POI Attributes and Descriptions (Q7)}
Attribute and description filters let users ask for stops that plain categories cannot express, such as a wheelchair-accessible supermarket or a museum about local history. This set of 200 queries that are based on Q4 (Ordered POI) that adds additional constraints on other POI attributes including the description. This tests the effectiveness of the description embeddings of Section~\ref{sec:filtering}. Each stop carries either a structured attribute, matched against the POI table, or a free-text description, matched by the embedding and LLM check of Section~\ref{sec:filtering}. We evaluate with the same pipeline as the main benchmark; \autoref{tab:desc} breaks down the outcomes per model.
\begin{table}[t]
\centering
\caption{Attribute- and description-filtered queries (Q7).}
\label{tab:desc}
\footnotesize
\setlength{\tabcolsep}{4pt}
\begin{tabular}{@{}l | rr r | rrr@{}}
\toprule
& \multicolumn{3}{c|}{\textbf{Correct}} & \multicolumn{3}{c}{\textbf{Incorrect}} \\
\cmidrule(lr){2-4} \cmidrule(l){5-7}
\textbf{Model} & Route & T-Inf & \textbf{Total} & Error & F-Inf & Viol \\
\midrule
Flash Lite  & \textbf{151} & 22 & \textbf{173} &  2 & \textbf{18} & \textbf{7} \\
GPT-4.1     & 114 & 20 & 134 & \textbf{1} & 51 & 14 \\
gpt-oss-20b &  99 & 19 & 118 & 16 & 24 & 42 \\
\bottomrule
\end{tabular}

\end{table}

\begin{figure}[t]
\centering
\begin{minipage}{\craftfigbox}
\begin{minipage}[b]{0.54\craftfigbox}
\centering
\begin{minipage}[c][81pt][c]{\linewidth}\centering\definecolor{cBlue}{HTML}{2C5F8A} 
\definecolor{cGray}{HTML}{7A7A7A} 
\begin{tikzpicture}
\begin{axis}[
    width=\linewidth, height=3.5cm,
    xlabel={Cosine threshold $\theta$}, ylabel={\% of POIs},
    xmin=0.20, xmax=0.70, ymin=0, ymax=108,
    xtick={0.2,0.3,0.4,0.5,0.6,0.7}, ytick={0,50,100},
    grid=major, grid style={gray!25},
    tick align=outside, tick pos=left,
    xlabel near ticks, ylabel near ticks,
    label style={font=\tiny}, tick label style={font=\tiny},
]
\draw[dashed, gray!70] (axis cs:0.32,0) -- (axis cs:0.32,108);
\addplot[color=cBlue, line width=0.9pt, mark=none] coordinates {
(0.20,99.71)(0.21,99.71)(0.22,99.71)(0.23,99.71)(0.24,99.42)(0.25,99.42)
(0.26,99.14)(0.27,99.14)(0.28,98.27)(0.29,97.98)(0.30,97.12)(0.31,95.97)
(0.32,94.81)(0.33,94.52)(0.34,93.66)(0.35,89.91)(0.36,85.88)(0.37,82.71)
(0.38,81.56)(0.39,79.25)(0.40,75.22)(0.41,72.05)(0.42,67.44)(0.43,61.96)
(0.44,57.64)(0.45,52.74)(0.46,51.30)(0.47,50.14)(0.48,48.42)(0.49,45.82)
(0.50,41.79)(0.51,40.92)(0.52,39.19)(0.53,27.67)(0.54,25.36)(0.55,23.92)
(0.56,22.19)(0.57,21.04)(0.58,18.73)(0.59,17.58)(0.60,16.71)(0.61,16.14)
(0.62,15.27)(0.63,13.54)(0.64,13.54)(0.65,12.39)(0.66,11.82)(0.67,11.24)
(0.68,9.80)(0.69,8.93)(0.70,8.07)
};
\addplot[color=cGray, line width=0.9pt, mark=none] coordinates {
(0.20,88.94)(0.21,88.16)(0.22,87.17)(0.23,86.12)(0.24,84.33)(0.25,82.54)
(0.26,80.71)(0.27,79.07)(0.28,76.95)(0.29,74.81)(0.30,72.33)(0.31,70.25)
(0.32,68.20)(0.33,65.78)(0.34,62.42)(0.35,59.82)(0.36,57.58)(0.37,54.68)
(0.38,51.72)(0.39,49.01)(0.40,46.05)(0.41,42.53)(0.42,38.54)(0.43,34.97)
(0.44,31.14)(0.45,27.92)(0.46,24.44)(0.47,21.56)(0.48,18.73)(0.49,15.78)
(0.50,13.44)(0.51,10.79)(0.52,8.71)(0.53,6.43)(0.54,4.63)(0.55,3.88)
(0.56,2.87)(0.57,2.28)(0.58,1.82)(0.59,1.22)(0.60,1.04)(0.61,0.79)
(0.62,0.62)(0.63,0.47)(0.64,0.45)(0.65,0.37)(0.66,0.36)(0.67,0.32)
(0.68,0.28)(0.69,0.26)(0.70,0.23)
};
\addplot[only marks, mark=*, mark size=1.1pt, color=cBlue] coordinates {(0.32,94.81)};
\addplot[only marks, mark=*, mark size=1.1pt, color=cGray] coordinates {(0.32,68.20)};
\node[anchor=west, font=\scriptsize, color=cBlue]   at (axis cs:0.44,101) {matches kept};
\node[anchor=west, font=\scriptsize, color=cGray] at (axis cs:0.44,82) {sent to LLM};
\node[anchor=west, font=\tiny] at (axis cs:0.205,18) {$\theta{=}0.32$};
\end{axis}
\end{tikzpicture}\end{minipage}\\[0.2em]
{\scriptsize (a) Choosing $\theta$}
\end{minipage}\hfill
\begin{minipage}[b]{0.44\craftfigbox}
\centering
\begin{minipage}[c][81pt][c]{\linewidth}\centering\footnotesize
\setlength{\tabcolsep}{1pt}
\begin{tabular}{@{}l rrr@{}}
\toprule
 & \textbf{Prec.} & \textbf{Rec.} & \textbf{Calls} \\
\midrule
LLM only         & 84.1 &  97.4 & 1.00 \\
Emb only         & 43.9 & 100.0 & 0.00 \\
\textbf{Emb+LLM} & \textbf{84.1} & \textbf{97.4} & \textbf{0.83} \\
\bottomrule
\end{tabular}
\end{minipage}\\[0.2em]
{\scriptsize (b) Cost per stop}
\end{minipage}
\end{minipage}
\caption{The description check. (a)~Choosing $\theta$ on the calibration phrases,
with the loss fixed at 5\% in advance. (b)~Each arm over the 47 description
stops of Q7; calls are relative to LLM only.}
\label{fig:threshold}
\end{figure}

\begin{table}[t]
\centering
\caption{Embedding quality. AUC and true matches at
$\theta = 0.32$ (\%).}
\label{tab:cs-phrase}
\footnotesize
\setlength{\tabcolsep}{5pt}
\begin{tabular}{@{}llrr@{}}
\toprule
\textbf{Phrase} & \textbf{Role} & \textbf{AUC} & \textbf{Kept at $\theta$} \\
\midrule
an art museum                      & test  & 0.91 & 100.0 \\
a history museum                   & test  & 0.89 & 100.0 \\
a theme park attraction            & test  & 0.85 &  90.4 \\
an upscale fine-dining restaurant  & calib.& 0.81 &  97.2 \\
a cafe that roasts its own coffee  & calib.& 0.80 & 100.0 \\
a park with sports fields          & calib.& 0.65 &  95.2 \\
a park with hiking trails          & calib.& 0.44 &  78.8 \\
\bottomrule
\end{tabular}
\end{table}

\noindent\textbf{Candidate-stop filter.}
Description filters are expensive since each candidate POI may need an LLM call. The embedding step cuts those calls by dropping POIs whose cosine similarity to the phrase is below $\theta$, and the LLM checks the rest.

We pick $\theta$ on about 12{,}000 phrase--POI pairs
covering every described restaurant, cafe, and park under four phrases no
benchmark query uses. GPT-4.1 labels each pair in two passes with the order
reversed, and pairs the two passes disagree on are dropped. Dropping a match is
the one error we cannot undo, since the LLM never sees it, while keeping a
non-match costs only one extra call. Fixing that loss at 5\% in advance gives
$\theta = 0.32$ (\autoref{fig:threshold}a). On the three benchmark phrases, which
took no part in this, the same $\theta$ keeps 93.2\% of the matches. The embedding
never accepts a POI on its own, since its quality swings by phrase
(\autoref{tab:cs-phrase}).

\autoref{fig:threshold}b measures the check inside the search. The Q7 queries
carry 47 description stops between them: calls are LLM requests, one per POI
checked, counted over all 47 and reported relative to the LLM-only arm, while
precision and recall cover the 39 stops whose phrase has labels. The embedding
alone keeps every true match but reaches only 43.9\% precision, so it cannot
decide on its own. The LLM alone is accurate but costs the most calls. CRAFT
runs both, which keeps the LLM's precision and recall while cutting the calls
to $0.83$, so the embedding pays for itself in work avoided rather than in
accuracy.
\smallskip
\subsection{Scalability with Long Trips (Q1,Q8)}
\label{sec:scalability}

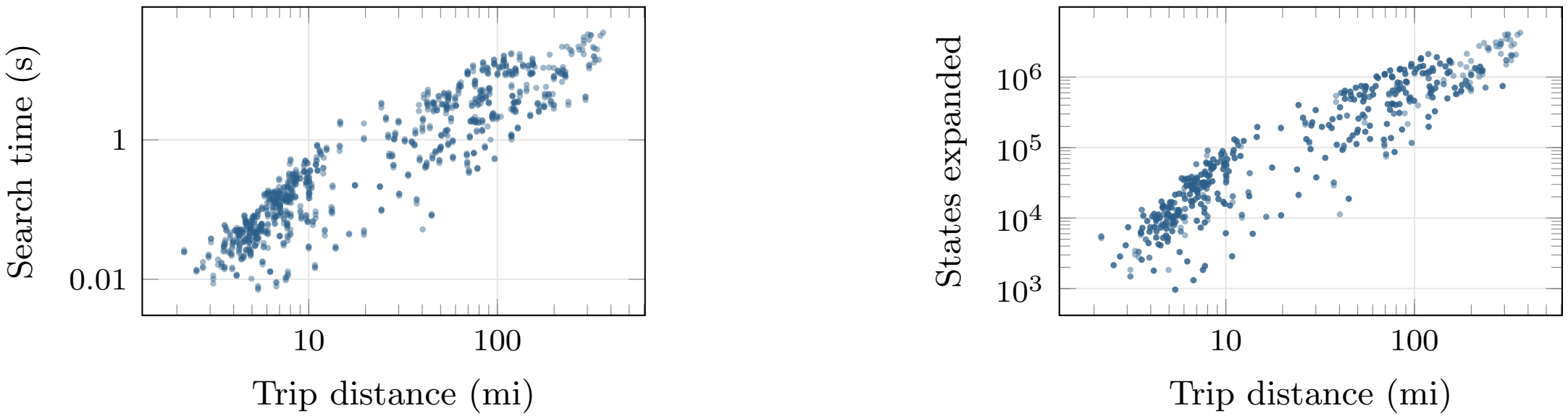
\begin{figure}[t]
\centering
\definecolor{cPts}{HTML}{2C5F8A} 
\pgfplotsset{
  scatterstyle/.style={
    width=0.52\craftfigbox, height=4.4cm,
    xlabel={\footnotesize Trip distance (mi)},
    tick label style={font=\scriptsize},
    label style={font=\footnotesize},
    grid=major, grid style={draw=black!10}
  }}
\begin{tikzpicture}
\begin{axis}[scatterstyle,
    xmode=log, ymode=log,
    log ticks with fixed point,
    ylabel={\footnotesize Search time (s)}]
\addplot[only marks, mark=*, mark size=0.6pt, color=cPts, opacity=0.45]
  table[col sep=comma, header=has colnames, x index=0, y index=1] {charts/craft_scaling.csv};
\end{axis}
\end{tikzpicture}%
\hfill
\begin{tikzpicture}
\begin{axis}[scatterstyle,
    xmode=log, ymode=log,
    log ticks with fixed point,
    ytick={1e3,1e4,1e5,1e6},
    yticklabels={$10^3$,$10^4$,$10^5$,$10^6$},
    ylabel={\footnotesize States expanded}]
\addplot[only marks, mark=*, mark size=0.6pt, color=cPts, opacity=0.45]
  table[col sep=comma, header=has colnames, x index=0, y index=2] {charts/craft_scaling.csv};
\end{axis}
\end{tikzpicture}
\caption{CRAFT search cost versus trip distance on EV Routing.}
\label{fig:scaling}
\end{figure}
\begin{table}[t]
\centering
\caption{CRAFT on the 200 long EV trips: the baseline engine (Long) vs.\ the same generated code under an A$^\star$ engine (Opt).}
\label{tab:shortlong}
\footnotesize
\setlength{\tabcolsep}{4pt}
\begin{tabular}{@{}l | rr | rr | rr@{}}
\toprule
& \multicolumn{2}{c|}{Success (\%)} & \multicolumn{2}{c|}{Search (s), med.} & \multicolumn{2}{c}{States, med.} \\
\textbf{Model} & Long & Opt & Long & Opt & Long & Opt \\
\midrule
Flash Lite  & 94.5 & 96.0 & 3.6 & 0.6 & 678k & 102k \\
GPT-4.1     & 75.0 & 92.5 & 3.1 & 0.5 & 584k & 101k \\
gpt-oss-20b & 71.6 & 87.3 & 3.1 & 0.5 & 594k &  98k \\
\bottomrule
\end{tabular}
\end{table}

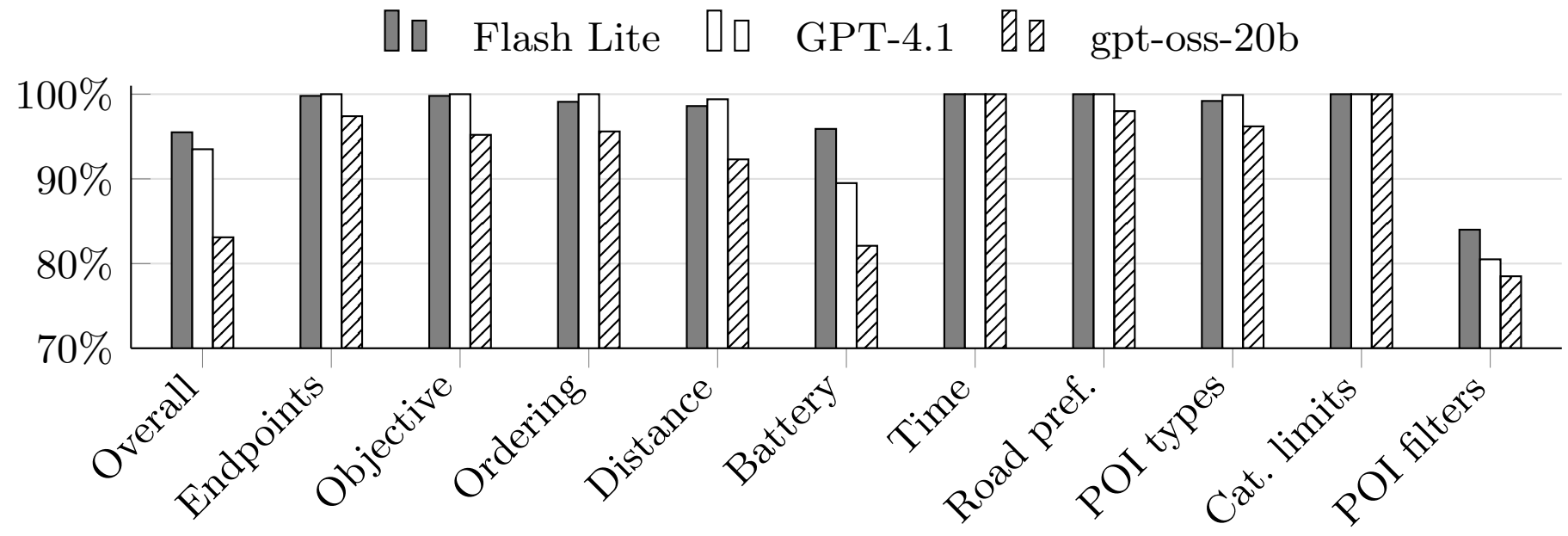
\begin{figure}[t]
\centering
\begin{tikzpicture}
\begin{axis}[
    ybar,
    width=1.06\craftfigbox,
    height=3.6cm,
    bar width=4.5pt,
    bar shift auto=0pt,
    enlarge x limits={abs=0.55cm},
    ymin=70, ymax=101,
    ytick={70,80,90,100},
    yticklabel={\pgfmathprintnumber{\tick}\%},
    tick label style={font=\footnotesize},
    ymajorgrids, grid style={draw=black!12},
    axis x line*=bottom, axis y line*=left,
    symbolic x coords={Overall,Endpoints,Objective,Ordering,Distance,Battery,Time,Road pref.,POI types,Cat.\ limits,POI filters},
    xtick=data,
    xticklabel style={rotate=45, anchor=east, font=\footnotesize},
    legend style={at={(0.5,1.03)}, anchor=south, legend columns=-1,
        draw=none, font=\footnotesize, column sep=8pt},
]
\addplot[fill=gray, draw=black] coordinates {
 (Overall,95.5) (Endpoints,99.8) (Objective,99.8) (Ordering,99.1) (Distance,98.6)
 (Battery,95.9) (Time,100) (Road pref.,100) (POI types,99.2) (Cat.\ limits,100) (POI filters,84)};
\addplot[fill=white, draw=black] coordinates {
 (Overall,93.5) (Endpoints,100) (Objective,100) (Ordering,100) (Distance,99.4)
 (Battery,89.5) (Time,100) (Road pref.,100) (POI types,99.9) (Cat.\ limits,100) (POI filters,80.5)};
\addplot[pattern=north east lines, pattern color=black, draw=black] coordinates {
 (Overall,83.1) (Endpoints,97.4) (Objective,95.2) (Ordering,95.6) (Distance,92.3)
 (Battery,82.1) (Time,100) (Road pref.,98) (POI types,96.2) (Cat.\ limits,100) (POI filters,78.5)};
\legend{Flash Lite, GPT-4.1, gpt-oss-20b}
\end{axis}
\end{tikzpicture}
\caption{Spec Builder field accuracy per model.}
\label{fig:parser-eval}
\end{figure}

This experiment tests how CRAFT's search scales with graph size. We extend (Q1) with 200 additional queries on the statewide network, whose trips are far longer than the city ones. Figure~\ref{fig:scaling} shows the relationship between the trip distance and the total search time (on the left) and the total number of expanded states during the search (on the right) for all 400 queries in both sets (Q1\&Q8). Both graphs consistently show that longer trips take more time. It also illustrates the direct correlation between the number of expanded states and the search time which allows us to measure the cost directly in terms of number of expanded states.


Because the generated functions are solver-agnostic, the solver itself can be optimized without generating new code. In this part, we use an optimized routing solver based on A$^\star$ which only changes the priority of expanding states to the Euclidean distance to destination. \autoref{tab:shortlong} compares the two engines. The jump in success rate is not the \underline{opt}imized engine finding routes the baseline could not: it is the same routes found sooner. Expanding states toward the destination cuts the median search from about 3~seconds to about 0.5, so queries that had been hitting the 900~second limit now finish inside it. Every one of the 69 queries it gains was a baseline timeout, no query is lost, and the 48 queries that have no route at all are the same under both engines. The gain is therefore largest where timeouts were: GPT-4.1 gains 35 and gpt-oss-20b 31, while Flash Lite, already finishing almost everything, gains 3.

\subsection{Spec Builder Accuracy}

We evaluate the Spec Builder stage in isolation on all of our queries, comparing every extracted field against the ground truth. Figure~\ref{fig:parser-eval} reports per-field accuracy. Endpoints and objective are near-perfect for every model, and most constraint fields stay above 92\%. The weakest fields are battery parameters and free-text POI filters, where qualitative phrases map imperfectly onto schema attributes. Overall, Flash Lite produces a fully correct structured query for 95.5\% of inputs, GPT-4.1 for 93.5\%, and gpt-oss-20b for 83.1\%, so parsing is rarely the bottleneck.

\section{Conclusion and Future Work}
\label{sec:conclusion}
We presented CRAFT, a system that answers constrained routing queries written in plain language. An LLM turns each query into the four functions a fixed solver needs, so one engine handles query types that would otherwise each need their own algorithm, and it is more accurate than both baselines with every backend we tested. Today every requirement is a hard constraint: a route either meets it or is thrown away, and all of them count the same. In practice some matter far more than others, since running out of battery is worse than missing one of the requested stops. We want the solver to rank requirements instead of treating them as equally binding, taking the order from the user when they give one and from sensible defaults when they do not, so that when no route satisfies everything it can relax the least important requirement first and return a few routes that give up different things. We also want to feed generation errors back to the LLM and retry, since bad generated code causes most of what still fails.

\bibliographystyle{IEEEtran}
\bibliography{references}

\end{document}